\documentstyle[aasms4]{article}
\begin {document}
\baselineskip=12pt

\title{ Flux Expulsion -- Field Evolution in Neutron Stars}
\author{M. Jahan-Miri}
\affil{Institute for Advanced Studies in Basic Sciences, Zanjan, 45195,
IRAN}
\authoremail{jahan@iasbs.ac.ir}

\begin{abstract}
Models for the evolution of magnetic fields of neutron stars are
constructed, assuming the field is embedded in the proton superconducting
core of the star. The rate of expulsion of the magnetic flux out
of the core, or equivalently the velocity of outward motion of
flux-carrying proton-vortices is determined from a solution of
the Magnus equation of motion for
these vortices.  A force due to the pinning interaction between
the proton-vortices and the neutron-superfluid vortices is also
taken into account in addition to
the other more conventional forces acting on the proton-vortices.
Alternative models for the field evolution are considered based
on the different possibilities discussed for the effective values of
the various forces.  The coupled spin and magnetic evolution of
single pulsars as well as those processed in low-mass binary
systems are computed, for each of the models. The predicted 
lifetimes of active pulsars, field strengths of the very old
neutron stars, and distribution of the magnetic fields versus
orbital periods in low-mass binary pulsars are used to test the
adopted field decay models. Contrary to the earlier claims, the
buoyancy is argued to be the dominant driving cause of the flux
expulsion, for the single as well as the binary neutron stars.
However, the pinning is also found to play a crucial role which
is necessary to account for the observed low field binary and
millisecond pulsars.
\end{abstract}


\section{INTRODUCTION}  
The observed high rate of incidence of binary pulsars among the
population of low-magnetic field pulsars indicates a field decay
mechanism in a neutron star which is linked with its evolution in a
binary system (eg., Phinney \& Kulkarni
1994).  This is also supported by even lower field strengths of the
millisecond pulsars which are generally believed to be the end
products of the binary evolution of pulsars in systems with low-mass
main-sequence companions (eg., van den Heuvel 1994).  A model for
the evolution of the magnetic field in neutron stars has been
proposed which indeed relates it to the spin evolution of the star,
in an intimate way. The model predicts that the magnetic flux is 
transported out of the core, at the same rate as that of the
spin-down of the star, into the crust where it 
undergoes Ohmic decay subsequently (Srinivasan~et~al.  1990). 
This field decay mechanism has been already tested
in a study of the spin evolution of pulsars in binary systems. 
The distribution of the final field strengths against the orbital
periods was found to be in good agreement with that of the observed
wide-orbit low-mass binary pulsars (Jahan-Miri \& Bhattacharya 1994). 
Also, implications of the model for the evolution
of single pulsars as well as those processed in massive binaries 
were shown to be consistent with, and provide new insights into, the
existing observational data on the radio and X-ray pulsars
(Jahan-Miri 1996a).

The above flux expulsion model (the so-called model of spin-down-induced
flux expulsion, hereafter the SIF model) assumes that,
as the star spins down, the outward moving (neutron) vortices
pull the fluxoids (the proton-vortices) along with them because of
their interpinning. However, given the finite strength of the pinning
interaction energy and also the presence of resistance against the motion
of both the fluxoids and the vortices the two might be, in general,
expected to move with different radial velocities, while ``cutting''
or ``creeping'' through each other.
A more refined treatment of the flux expulsion, therefore, requires 
the dynamics of fluxoid motion to be considered explicitly, as is
attempted here. A further improvement, not addressed here, would be
to solve simultaneously for the dynamics of the fluxoids and
the vortices, self-consistently. Our treatment is, however, 
justified for the steady-state spin-down periods when the core
superfluid spins down at a given rate which determines the radial
velocity of the vortices. 

In the following, we discuss the various forces which act on the
fluxoids in the interior of a neutron star, including {\em i)} a
force due to their pinning interaction with the moving vortices.
The other forces which we take into account are {\em ii)} viscous
drag force due to magnetic scattering of electrons, {\em iii)}
buoyancy force, and {\em iv)} curvature force.  
The velocity of the outward motion of the fluxoids could be then 
determined, for a given steady-state spin-down rate of the star,
from a solution of the Magnus equation requiring a balance of the
forces acting radially on a fluxoid.
The derived radial velocity of the fluxoids at the core-crust
boundary would in turn determine the rate of the flux expulsion out
of the core. Our approach here is analogous to that of Ding, Cheng \&
Chau (1993, hereafter DC93) in
their study of the field evolution of single normal pulsars (see
also Muslimov \& Tsygan 1985; Harvey, Ruderman \& Shaham 1986;
Jones 1987, 1991).  The original motivation for the present work
was to investigate the results of such a treatment of the
flux expulsion scenario for the magnetic evolution of {\em binary
pulsars} for which the spin-down history of the neutron star is
quite different than that of the single pulsars
addressed in DC93.  Furthermore, we have also explored
other possibilities different than those assumed by DC93
about the nature and magnitude of the various forces
acting on the fluxoids. The resulting alternative models for
calculating the rate of flux expulsion could, in principle, probe
the physics of the interior of neutron stars in more details.
Predictions of these models for the magnetic evolution
of binary as well as single pulsars are compared and tested
against the available observational data.

\section{Dynamics of Fluxoids}
\subsection{The Scene}
Superconductivity of the protons in the interior of a neutron
star and the presence of a lattice of fluxoids carrying the 
magnetic flux in the core form the underlying premise for
the field decay mechanism under study.
As it is the case also with the assumed superfluidity of neutrons in a
neutron star, the protons are speculated to form a (type II)
superconductor based on the microscopic calculations of the associated
energy gap in the BCS theory (Ruderman 1972).
The corresponding
transition temperature for the proton-superconductivity,
though smaller than that of the neutron-superfluidity in
the crust, is similar to that of the neutron-superfluidity
in the core (se, eg., Sauls 1989) and much larger than the
expected interior temperatures for the observed pulsars
(Pines \& Alpar 1992).
Doubts against superconductivity of the protons have
been, however, raised because the earlier calculated
core coupling time scale (Alpar \& Sauls 1988), for the case of
superconductor protons, could not successfully accommodate 
the observed post-glitch recovery of the Vela pulsar (Pines \& Alpar
1992). This difficulty does not exists anymore since a
necessary correction
in the calculation of the coupling time scale results in more than
an order of magnitude reduction in its value and makes it much
smaller than the inferred observational upper limit
(Jahan-Miri 1998). Moreover, the velocity relaxation time
scale of the vortices for the case of normal protons in the core
would be much larger than, or at the best (for the very cool neutron
stars) similar to, the case of superconducting protons (see, eg.,
Sauls 1989) . Hence
normal protons may not offer any solution to the problem. On the other
hand, the additional coupling mechanism due to an interpinning of the
fluxoids and the vortices could result in a reduction in the velocity
relaxation time of the vortices as well as in the coupling time
for the superfluid bulk matter (estimates of the predicted time
scales may be found in Jahan-Miri 1998). Thus superconductivity
of the protons is rather favored by the above observational
restriction on the core coupling time scale.

Given a superfluid-superconductor mixture in the interior of a
neutron star with the associated (neutron) vortices
and (proton) fluxoids, we further assume that both families of the
vortex lines form regular uniform lattices parallel to the spin and
magnetic axes, respectively.  A stabilizing toroidal field component
might be also present in the core fluid, should
the crystalization of the crust does not take place early enough
after the birth of a neutron star (Flowers \& Ruderman 1977).
The toroidal component, if it is assumed to be ($>> 10^{15}$~G)
much larger than the observable dipolar field would quench the
superconductivity of the protons in the core, altogether. 
Otherwise, and if it is frozen, along with the dipolar field,
within the superconducting core it would
result in a complicated distorted geometry of the fluxoids.
Such ditorted geometries of lines, which is neglected here
for simplicity, would affect our following treatment of the 
flux expulsion in two ways. The effective total length of
the fluxoids would be more than in the case of a regular
parallel lattice assumed here. And, the entanglement
of the fluxoids with the neutron vortices would be realized
more effectively even in the cases when the rotation and
magnetic axes of the star are nearlly parallel. The case of
nearly parallel axes are indeed preferred in our models (see
the next paragraph) thus the introduction of a distorted
fluxoid geometry would have points for and against the
following analysis. 
The rate of the Ohmic decay of the expelled flux in the crust
is subject to uncertainties about the correct value of the
conductivity of the crust matter as well as the geometry and 
transport behavior of the expelled flux due to other possible
processes, namely the Hall drift and the turbulent cascade mechanism
(see, eg., Jones 1988; Goldreich \& Reisenegger 1992; Bhattacharya \&
Datta 1996). Different values for the effective decay time scale of
the field in the crust are therefore tested in our model computations,
to bypass these complications.

Throughout, we will be considering the motion of the fluxoids only in
the region close to the core-crust boundary that also lies in the
magnetic equatorial plane.  The strength of the
average field of the stellar core is decreased by the transport of
these boundary fluxoids out of the core; the fluxoids in the interior
regions are assumed to adjust their positions accordingly to maintain
a uniform density. We will be referring to a cylindrical coordinate
system aligned with the fluxoids in considering the radial and
azimuthal directions, which coincide with their counterparts in
a spherical system for the above region of interest. The radial
velocities of the neutron vortices, in the same region,
would however be co-linear with that of the fluxoids only for the
vortex segments located near the spin equator of the star as well.
The fractional
size of such a region (of coincidence of the magnetic and spin
equators) would be larger, hence our treatment of the radial
velocities and forces for the fluxoids and the vortices would be
more accurate, for the smaller inclination angles.

It has been suggested (Sauls 1989; Jones 1991) that, since the
pinning interaction energy is independent of a displacement of
a pinned neutron vortex parallel to itself, the vortices
might be able to slide along the fluxoids without producing a
large scale movement of the fluxoids.  It is noted, however, that
such a sliding might be realized only for the vortices in some
parts of the spin equator, namely for those lying at large
magnetic latitudes. Moreover, this possibility does not by itself
violate our following assumption of a
radial reactive force acting on the fluxoids even in the regions
where sliding might occur.  The Magnus force on a sliding vortex
would have the same magnitude and direction as for the non-sliding
vortices and its component in the direction perpendicular to the
fluxoid has to be balanced in any case with a force due to the
pinning which will exert a reaction force on the fluxoid.
Furthermore, since the assumed sliding cannot be realized for all of
the vortices, for any assumed geometry of the lines, its partial
occurrence, if at all, seems to be further questionable on the
account that it would result in an azimuthally non-uniform
distribution of the vortex density.

\subsection{Neutron Superfluid Vortices} 
In the steady-state, the vortices in the core of a neutron star
are expected to be co-rotating with the charged component of the
star, including the lattice of the proton fluxoids, at a given
rate $\Omega$ (Sauls 1989). However, the superfluid bulk matter
has to rotate faster at a rate $\Omega_{\rm s}$, maintaining a
rotational lag \( \omega \equiv \Omega_{\rm s} - \Omega \) with
its vortices, in
order to follow the spinning down of the star at a rate
$\dot \Omega$. The corresponding radial velocity $v_{\rm n}$ of
the vortex outward motion, at the core boundary, would be 
\begin{eqnarray}
v_{\rm n} &=& -{ R_{\rm c} \over 2} { {\dot \Omega } \over
\Omega_{\rm s} } \approx -{ R_{\rm c} \over 2} { {\dot \Omega}
\over \Omega } \nonumber \\ 
&=& 1.59 \times 10^{-2} { \dot P_{\rm
yr} \over P_{\rm s} } \ \ {\rm cm \ s}^{-1} 
\end{eqnarray} 
where
$R_{\rm c} = 9 \times 10^5$~cm is the assumed radius of the core
of a neutron star, and $P_{\rm s}$ and $\dot P_{\rm yr}$ are the
spin period and its time derivative in units of s and ${\rm s \
yr}^{-1}$, respectively. The superfluid spins down as the number
of vortices is decreased due to their outward motion. The
rotational lag $\omega >0$, in a spinning down superfluid, results
in an outward radial Magnus force $F_{\rm M} = \rho_{\rm s}
\kappa R_{\rm c} \omega$ on the vortices, per unit length at
the core boundary, where $\rho_{\rm s}$
is the neutron-superfluid density, and
$\kappa= 2 \times 10^{-3} {\rm cm}^2 {\rm s}^{-1}$ is the vorticity
of a vortex line. The Magnus force is usually assumed to be
balanced by the viscous force against the motion of the vortices
which is primarily caused by the scattering of electrons off the
magnetized cores of neutron vortices in the interior of a neutron
star (Sauls 1989).

\vspace{.5 cm} 
\noindent 
{\bf The pinning force on the neutron vortices :}\\ 
On the other hand, if a pinning interaction is associated with
an intersecting fluxoid-vortex pair the corresponding pinning force
on the vortex could be, in principle, in either (inward or outward)
radial directions. The force direction is decided by the direction
of the relative motion of the interacting vortex and fluxoid, and is
independent of the actual direction of motion of the vortices. That
is, the pinning force could contribute to the viscous forces against
the outward motion of the vortices if they move faster than the
fluxoids. However, in the opposite case it would act as a
``driving'' force for the spinning down of the superfluid, being in
the same direction as the outward motion of the vortices. Moreover,
the viscous force due to the electron scattering is expected to be
many orders of magnitudes smaller than the pinning force, even for
the largest spin-down rates of interest and hence for the largest
possible values of $v_{\rm n}$. Typically, the pinning force, per
unit length of a vortex, is expected to be $>> 10^{12}$~dyn~cm$^{-1}$,
while the viscous force of the electron scattering (Alpar \& Sauls 1988) 
is $< 10^5$~dyn~cm$^{-1}$ (and the buoyancy force on the magnetized
neutron-vortices is even smaller than this). Therefore, for the
vortices the pinning forces exerted by the fluxoids have to be
balanced by the Magnus force on them.

An interesting consequence of such a balance of forces for the
vortices is that the superfluid might be rotating {\em slower} than its 
vortices ($ \omega < 0 $) while it is {\em spinning down}. This is in
contrast to the usual conditions during a spin-down phase of a
superfluid, in general and, particularly when pinning is present and
larger values of the lag are the case. As indicated, a spinning down
superfluid
must be rotating faster than its vortices ($ \omega > 0 $),
so that the viscous force (as well as the pinning force if it is
present) on the outward moving vortices is balanced by an outward
driving Magnus force. However, since for the above assumed
case the pinning force is itself directed {\em outward} it could
as well play the role of the driving force, which in turn
requires the balancing Magnus force to be directed inward. Thus,
a negative rotational lag \hbox{ ($ \omega = \Omega_{\rm s} -
\Omega< 0 $)}, hence an inward Magnus force, during a spin-down
phase of the superfluid in the core of a neutron star is realized
when the fluxoids move {\em faster} than the vortices.

\subsection{The Pinning Force on the Fluxoids} 
In any case, a pinning force of the same magnitude as, and in the
opposite direction to, that exerted {\em by} the fluxoids will be
also exerted by the vortices {\em on} the fluxoids. Considering
the above discussed balance of the forces on vortices, and
equating the forces communicated between the two lattices of
the vortices and the fluxoids, per unit volume, the pinning
force $F_{\rm n}$ acting on a fluxoid, per unit length, is
derived as
\begin{eqnarray}
F_{\rm n} &=& { n_{\rm v} \over n_{\rm f}} F_{\rm M} \approx 
	2 \phi_{0} \ \rho_{\rm s} \ R_{\rm c} \ 
	{\Omega(t) \ \omega(t) \over B_{\rm c}(t)} \nonumber \\
	&=& 5.03  { \omega_{-6} \over P_{\rm s} B_{8}} \ \  
	{\rm dyn cm}^{-1}
\end{eqnarray}
where $n_{\rm v} = { 2 \Omega_{\rm s} \over \kappa}$ and $n_{\rm f}=
{B_{\rm c} \over \phi_{0}}$ are the number densities per unit cross
section area of the vortices and the fluxoids, respectively,
$\phi_{0}=2 \times 10^{-7} {\rm G\ cm}^2$ is the
magnetic flux carried by a fluxoid, $B_{\rm c} = 10^8 B_{8}$ is the
strength of the core field in units of G, and $\omega_{-6}$ is the
superfluid lag $\omega$ in units of $10^{-6} {\rm rad \ s}^{-1}$.
Notice that the sign of $\omega$ determines the sign of $F_{\rm n}$
for which, as well as for the other forces discussed below, the
outward direction will be reckoned as the positive sense.

The above derivation of $F_{\rm n}$ assumes that the total Magnus
force acting {\em continuously on all} the vortices is 
communicated instantaneously to {\em all} the fluxoids. In contrast, 
it might be argued that in general only a small
fraction of the vortices would be directly interacting with the
fluxoids at any instant
of time.  The remaining much greater fraction of them (of the order
of the ratio of an inter-fluxoid spacing to the size of a pinning
interaction region) should reside in the inter-fluxoid
spacings.  Also, the assumed total force on the fluxoids has been
divided equally among all in spite of the similar expectation that at any
given time a majority of them would be located in the inter-vortex
regions, far from any pinning site.  Nevertheless, the motion of the
fluxoids (as well as the vortices) is further constrained due to
their mutual repulsive forces, which would require a uniform density
of the lines to be maintained, in a steady state.  Consequently, the 
fluxoids (whether being in an interaction region or in a free
region) are forced to move always together (on scales of, at least, 
an inter-vortex spacing) which requires that the force acting on
some of them to be shared equally among all, instantaneously, as
is assumed in Eq.~2.
The same argument fails however for the vortices since their
displacements on scales of the order of an inter-fluxoid spacing
(which is many orders of magnitudes smaller than the inter-vortex
spacing) is not prohibited by the above requirement of having a
uniform density. On the other hand, and for the same reason, any
assumed vortex which is not interacting with a fluxoid at a
given time is expected to move relative to the rest of the
neighboring vortices and rapidly adjust its position (within a
distance of an inter-fluxoid spacing) until it also is located 
within a pinning interaction zone. That is all the vortices
would be continuously interacting with the fluxoids, which
completes the justification of the above derivation of $F_{\rm n}$.
Hence, Eq.~2 is valid during a vortex-fluxoid {\em co-moving}
phase which was (implicitly) assumed in the above arguments.

However, should the (radial) velocities of vortices and fluxoids be 
different, the above restriction on the vortex positions might not be
fulfilled steadily since, in this case, they have to continually
move through the inter-fluxoids spacings as well.  The effective
instantaneous force per unit length of a fluxoid, $F_{\rm n}$,
during such a state would be smaller than that given in Eq.~2, 
and could be estimated as the time averaged force acting on the
fluxoids (or equivalently by using the fractional number of the 
vortices interacting with the fluxoids at any time). Therefore,
during a phase of {\em unequal} velocities of the vortices and
fluxoids one derives (in contrast to Eq.~2 for a co-moving phase) 
\begin{eqnarray}
F_{\rm n} &=& { d_{\rm P} \over d_{\rm f}} \left( { n_{\rm v} \over 
	n_{\rm f}} F_{\rm M} \right)  \nonumber \\
        &=&2.59 \times 10^{-4}  { \omega_{-6} \over 
          P_{\rm s} B_{8}^{1/2}} \ \ {\rm dyn \ cm}^{-1}
\end{eqnarray}
where $d_{\rm f} = 2.3 \times 10^{-7} B_{8}^{-{1\over2}} $~cm is
the inter-fluxoid spacing, and $d_{\rm P}$ is the effective size of a
pinning interaction region around each fluxoid. A value of $d_{\rm P} =
\lambda_{\rm p}^* = 118$~fm has been used for the assumed
magnetic pinning mechanism (see below), where $\lambda_{\rm p}^* $
is the effective London length of the proton superconductor, being 
also a length scale for the spread of the magnetic field
of a neutron vortex line (Sauls 1989).

The ``averaged'' value of $F_{\rm n}$ as given in Eq.~3 assumes
that the velocity of a vortex while it is crossing through an
interaction region is the same as in the interaction-free space
between the fluxoids; equal weights have
been accordingly assigned to the corresponding time periods.
Realistically, however, the vortices might be expected to move much
faster in the free regions than in the pinning
regions because of the large difference in the effective resisting 
forces acting on them in the two environments (as discussed above).
As a consequence, they might tend to spend most of their time within the
pinning zones which would require an almost zero weight to be
assigned to the crossing times of the inter-fluxoid spacings. That is,
their dynamics would resemble the case of the co-moving phase and,  
$F_{\rm n}$ as in Eq.~2 would be applicable even when there is a
relative motion between the fluxoids and the vortices. 

Thus, while for a co-moving phase $F_{\rm n}$ is uniquely
determined by Eq.~2, when velocities are different one is faced
with two different possibilities as given by Eq.~2 and Eq.~3.
The two derivations represent, in fact, two extreme possibilities 
regarding the relative behavior of the different pinned segments of
a vortex during its motion. If each pinned segment of a vortex
could creep independently (over the length scales of about an
inter-fluxoid spacing) then the above argument to justify Eq.~2
applies to the phase of nonequal velocities as well. In contrast,
for a vortex
line of infinite rigidity the whole line moves always as a single
piece and $F_{\rm n}$ as in Eq.~3 would be appropriate. 
We will test both possibilities in the alternative models
which we use, in order to distinguish the relevant approximation
for the behavior of the vortices in the core of neutron stars.

\vspace{.5 cm}
\noindent
{\bf The `` critical lag'':} \\ 
The magnitude of the force which could be exerted {\em at each
intersection} by a vortex on a fluxoid, and vice versa,
is limited by a maximum value $f_{\rm P}$ corresponding
to the given strength of the pinning energy $E_{\rm P}$ and the
finite length scale of the interaction $d_{\rm P}$; namely
$E_{\rm P} = f_{\rm P} d_{\rm P}$.  The Magnus force on the vortices,
which has to be balanced by the pinning force, (and hence $F_{\rm n}$)
cannot therefore exceed a corresponding limit. This, in turn,
implies a maximum critical lag $\omega_{\rm cr}$, which is
determined by equating the Magnus force
\mbox{($=\rho_{\rm s} \kappa R_{\rm c} \omega$)} with the maximum
available pinning force \mbox{($=\frac{f_{\rm P}}{d_{\rm f}} \equiv
\frac{E_{\rm P}}{d_{\rm f} d_{\rm P}}$)}, per unit length of a
vortex. The pinning energy, in the magnetic interaction mechanism,
arises because of the difference in the free energies of a
fluxoid-vortex pair when they overlap at an intersection or are
separated. For a fluxoid with an average field $\vec{B}_{\rm p}$
and a vortex having an average field $\vec{B}_{\rm n}$, the magnetic
energy density in the overlapping case would include a 
term $(2/8 \pi) \vec{B}_{\rm p}.\vec{B}_{\rm n}$, in addition to
the sum of their individual contributions,
$(B^2_{\rm p} + B^2_{\rm n})/8 \pi$, when they are separated. The
pinning energy, per intersection, is therefore estimated
by multiplying the interaction overlap
volume $\sim (2 \lambda_{\rm p}) (\pi {\lambda^*_{\rm p}}^2) $ with
the above additional term of the energy density, 
which results in $E_{\rm P} \sim 10^{-5}$~ergs (see Jones 1991 for a
more refined derivation). The critical lag $\omega_{\rm cr}$ 
may be then calculated, as indicated above, to be 
\begin{eqnarray}
	\omega_{\rm cr} &=& 1.59 \times 10^{-6} \ B_{8}^{1/2} 
	\ \ {\rm rad \ s}^{-1}
\end{eqnarray}
where we have used the same parameter values as in DC93 in
order for the further comparison of the results. The critical lag
is the magnitude of the lag when velocities are different;
\hbox{ $ \omega = \omega_{\rm cr} $} or \hbox{ $ \omega = -
\omega_{\rm cr} $} when the vortices move faster or slower than
the fluxoids, respectively. However, during a co-moving phase
when the force communicated between a vortex and a
fluxoid at each pinning point is less than its maximum value,
$f_{\rm P}$, the lag might have any value within the range \hbox{ $
- \omega_{\rm cr} < \omega < \omega_{\rm cr}$}.

We note that a different estimate for the pinning energy due to the
proton density perturbation gives a smaller value of $E_{\rm P} \sim
5 \times 10^{-7}$~ergs.  The pinning energy in this case arises
because of a difference in the condensation energy between the pinned
and the free configurations, as a result of the change in the
proton density induced by the large velocity of the neutrons close
to the core of a neutron vortex (Sauls 1989).  The interaction volume
for this
mechanism is $\sim \xi^2_{\rm n} \xi_{\rm p}$, and the change in the
free energy density is estimated as \mbox{$n_{\rm n}
\frac{\Delta^2_{\rm p}}{E^2_{\rm F_{\rm p}}} \ \frac{\Delta^2_{\rm
n}}{E^2_{\rm F_{\rm n}}}$}, where $\Delta$ is the condensation
energy gap, $E_{\rm F}$ is the Fermi energy, $\xi$ is the coherence
length, $n$ is the number density, and the subscripts ``p'' and ``n''
refer to the protons and neutrons, respectively. The larger values of 
$E_{\rm P}$ in the case of the density perturbation is, however,  
associated with an interaction length, $d_{\rm P}=\xi_{\rm p}$, which is
smaller than for the magnetic interaction, $\lambda_{\rm p}^*$,
by about the same ratio as the inverse of the pinning energies.  
Thus, similar values of $f_{\rm P}$, and hence $ \omega_{\rm cr}$,
are expected for the both pinning mechanisms, as has been indicated
earlier (Bhattacharya and Srinivasan 1991).  Nevertheless, 
since the magnetic interaction depends on the angle between 
fluxoids and vortices the pinning caused by the density perturbation
might indeed have the dominant effect in a neutron star with a nearly
parallel geometry of the vortices and fluxoids.

\subsection{Other Forces on the Fluxoids}
In addition to the pinning force, the fluxoids in the interior of
a neutron star are also subject to the following radial forces.

\vspace{.5cm}
\noindent
{\bf The drag force:}\\ 
An isolated fluxoid moving through the normal degenerate electron
gas in the core of a neutron star is subject to the viscous drag force
of the electrons scattering off its magnetic field. The viscous drag
force, per unit length of a fluxoid, is estimated (Jones 1987) to be
\begin{eqnarray}
	\vec{F_v} &=& - {3 \pi \over 64} { n_{\rm e} e^2 \phi_0^2 
        \over E_{\rm F_{\rm e}} \lambda_{\rm p}} 
	{ \vec{v_{\rm p}} \over c}   \nonumber \\
	&=& -  7.30 \times 10^7 \ \vec{v_{\rm p}} \ \ 
	{\rm dyn \ cm}^{-1} 
\end{eqnarray}
where $v_{\rm p}$ is the velocity of the outward radial motion of
the fluxoids in units of ${\rm cm \ s}^{-1}$,  
$n_{\rm e} = 3.  \times 10^{36} {\rm cm}^{-3}$ is the number
density of the electrons, and $E_{\rm F_{\rm e}}= 88 $~MeV is the
electron Fermi energy, corresponding to a total density
$\rho = 2 \times 10^{14} {\rm g\ cm}^{-3}$, and a neutron number 
density $n_{\rm n} = 1.7 \times 10^{38} {\rm cm}^{-3}$ in the core.

The expression for $F_v$ in Eq.~5 is derived based on the
assumption of independent motions for single fluxoids. However,
for the typical conditions in the interior of a neutron star the
lattice of fluxoids might be, as a whole (or at least as bundles
consisting of not less than ten million fluxoids),
``frozen-in'' the electron gas (Harvey~et~al.  1986). This is
because the mean free path of the electrons turns out to be many
orders of magnitudes larger than the mean distance between their
successive scattering events, and also the deflection angle at each
event is very small. A treatment of the coherent electron scattering
by the fluxoid lattice has been shown to indeed require an almost
zero relative velocity between the electrons and the lattice (Jones
1987). Therefore, the flux expulsion out of the core might 
be prohibited except for the presence of electron-current loops
across the core-crust boundary. Uncertainties about the true
distribution of the magnetic flux and the correct value of the
conductivity in the crust, and also the possibility of a 
mechanical failure of the solid crust due to a build-up of
the magnetic stresses, however, obscure any definite conclusion 
to be drawn.

Moreover, there are other reasons to suspect the suggested
frozen-in approximation for the fluxoids as a whole (see also
DC93; Ruderman 1995), in the core of a neutron star. For example,
the finite volume of the fluxoid lattice and also the influence
of the superconductor boundary effects on the motion of the
fluxoids which have not been included in the previous studies of
the coherent scattering could as well have significant new 
consequences. In addition, the motion of the incompressible
electron fluid in the interior of a neutron star has been argued
to be divergence free (Goldreich \& Reisenegger (1992). The above 
permitted motion of the fluxoids, along with the electrons in the
frozen-in approximation, must be, therefore, of the same
(divergence free) nature. This is however impossible for the
uniform lattice of fluxoids during its outward motion since its
lattice constant keeps changing. Hence, a compromise between the
flux freezing and the divergence free motion of the electrons has
to be worked out if any flux expulsion is to be accounted for.

Given the above uncertainties as well as the lack of any other
definite prescription for calculating the drag force due to the
electron scattering, we use a value of $F_v$ as given by Eq.~5 
in our models. This choice is also supported by noticing that
a tentative expulsion time
scale derived for the case of coherent scattering, as implied by
the Hall drift of the flux in the base of the crust (Jones 1988,
1991), turns out to be similar to that based on
the single fluxoid approximation.
On the other hand, a maximum velocity for the outward motion
of the fluxoids has been suggested for the case of coherent
electron scattering (Jones 1991). Also, in some of our models
(models B1 and B2, described below) a similar
value for the maximum possible velocity of the fluxoids is
adopted, eventhough for a different reason.  The same models might
be therefore viewed also as models representing the frozen-in
approximation due to the coherent electron scattering.

\vspace{.5 cm}
\noindent
{\bf The buoyancy force:} \\
The buoyancy force on fluxoids in a neutron star arises for reasons
analogous to the case of macroscopic flux tubes in ordinary stars.
Because flux tubes are in pressure equilibrium with their 
surrounding the excess magnetic pressure causes a
deficit in the thermal pressure, and hence in the density, of the
plasma inside a flux tube which make the tube to become buoyant.
The radially outward buoyancy force $F_{\rm b}$ on a fluxoid,
per unit length, can be expressed as (Muslimov \& Tsygan 1985;
Jones 1987)
\begin{eqnarray}
	F_{\rm b} &=& \left( { \phi_{0} \over {4 \pi 
	\lambda_{\rm p}} }
        \right)^2 { \ln (\lambda_{\rm p} / \xi_{\rm p}) \over R_{\rm
	c}  }  \nonumber  \\
	  	  &=& 0.51 \ \ {\rm dyn \ cm}^{-1}
\end{eqnarray}
where values of $\lambda_{\rm p} = 131.5 $~fm and $ {\lambda_{\rm p}
\over \xi_{\rm p}} = \sqrt{2} $ have been used. 

Harrison (1991) has raised objection against the relevance of the
buoyancy force for the motion of single fluxoids in the core of
neutron stars, assuming
the {\em whole} lattice of the  fluxoids to be frozen in the electron-proton
plasma within the star. He argues that the buoyancy force would 
rather contribute to the gradient of
the macroscopic magnetic stresses supporting the hydrostatic
equilibrium of the plasma within the star, instead of acting on
the fluxoids individually. His argument is not, however, applicable 
if a relative motion between the fluxoids and the plasma is allowed
to take place, as he also makes it clear (see, in particular, the
last paragraph before \S~3.2, p.~422, in Harrison 1991). Thus, 
for our model calculations, having assumed a drag force corresponding
to the single fluxoid motion, we also take into account 
the buoyancy force on the fluxoids, for self-consistency. 
In this context, Ding et~al. (DC93) have argued that the above
Harrison's objection against the buoyancy force might be
irrelevant because the fluxoids motion could be fast enough
such that the
conditions of hydrostatic equilibium of the star is not
satisfied during their motion. Such an argument is
apparently missing the point, given that the independent
motions of single fluxoids has been also assumed in DC93.
The decision whether the buoyancy force acts on the fluxoid
lattice as a whole, hence contributing to the hydrostatic
pressure support, or else on each flux line individually is
independent of, and would be equally effective in, the
presence or absence of the equilibrium. 
Even if hydrostatic equilibrium within a neutron star is
assumed, as is speculated in DC93, to be approached on 
timescales $> 1$~Myr, corresponding to the largest fluxoid
velocities $< 10^{-7}$~cm~s$^{-1}$, the buoyancy force could
be still acting on the whole lattice, should a relative motion
between the fluxoids and the plasma being prohibited due to the
requirements of the electron scattering orbits off fluxoids.

\vspace{.5 cm}
\noindent
{\bf The curvature force :} \\
The tension of a vortex line (such as a fluxoid) implies that
a curved geometry of the line would result in a restoring force,
the curvature force, which tries to bring the line back to
its minimum energy straight configuration. The concavely directed
curvature force $F_{\rm c}$, per unit length, on a vortex having a
tension $T$ and a curvature radius $S$ is given as \(
F_{\rm c} = { T/ S } \) (Harvey~et~al.  1986). Thus, 
for a fluxoid in a neutron star, having a tension $T_{\rm p} = 
\left( { \phi_{0} \over {4 \pi \lambda_{\rm p}} }\right)^2 
{ \ln (\lambda_{\rm p} / \xi_{\rm p})}$, the magnitude of the curvature
force would be
\begin{eqnarray}  
\left| F_{\rm c} \right|  &=& {R_{\rm c} \over S} 
\left( { \phi_{0} \over {4 \pi \lambda_{\rm p}} }\right)^2 
{ \ln (\lambda_{\rm p} / \xi_{\rm p}) \over R_{\rm c}}  \nonumber  \\
         &\equiv& {R_{\rm c} \over S} F_{\rm b} 
\end{eqnarray}  
where we have used Eq.~6 to express the absolute value of $F_{\rm c}$
in terms of the (positive) buoyancy force. Moreover, the end
points of a fluxoid, where its magnetic flux spouts out and joins
the almost uniform field of the crust, are expected to be frozen
in at the bottom of the crust due to the large conductivity of
the matter.  An outward moving fluxoid might be therefore
expected to bend outward and becomes subject to an inward
curvature force, per unit length,  
\begin{eqnarray}
       F_{\rm c} &=& - {R_{\rm c} \over S_{\rm c}} F_{\rm b} \nonumber \\
                 &=& - 0.35  \ \ {\rm dyn \ cm}^{-1}
\end{eqnarray}
using Eq.~6, and a value of ${R_{\rm c} \over S} \sim \ln{2}$ for
the assumed spatially uniform distribution of the fluxoids (DC93).

The currents at the bottom of the crust are however subject to
diffusion processes, the rate of which would set a maximum
limiting speed $v_{\rm max}$ for the motion of the end points
of the fluxoids. We, therefore, assume that whenever
\( v_{\rm p} < v_{\rm max} \) the fluxoids remain straight and
{\em no} curvature force will be acting on them ($F_{\rm c}=0$),
since their end points are also able to move with the same speed.
In the opposite case, when \( v_{\rm p} \geq v_{\rm max} \), 
the fluxoids would be bent outward and the force
$F_{\rm c}$ as in Eq.~8 will be used.  We note that a
self-consistent solution of the equation of motion (Eq.~12 below)
requires further that during a transition
between the above two regimes $F_{\rm c}$ should change gradually
(see Fig.~3 below). We have allowed for such a smooth variation of
$F_{\rm c}$ at and around \( v_{\rm p} = v_{\rm max} \), in the
models which, otherwise, use the above prescription for calculating  
$F_{\rm c}$. The above maximum drift velocity $v_{\rm max}$ of
the magnetic flux in the crust is estimated, based on the Ohmic
diffusion alone, as
\begin{eqnarray}
        v_{\rm max} &\sim& {R \over \tau }   \nonumber   \\
        &=& 3.18 \times 10^{-9}  \left( { \tau \over
        10^7 {\rm yr} } \right)^{-1}  \ \ {\rm cm \ s}^{-1}
\end{eqnarray}
where $R=10^6$~cm  is the radius of a neutron star, and 
$\tau $ is the assumed time scale, in units of yr,
for the decay of the magnetic field in the crust. A larger value
for $v_{\rm max}$ may be expected if the Hall drift of the magnetic
flux at the bottom of the crust is also taken into account.
Also, the mechanical failure, and the plate tectonic
motion, of the crust (Ruderman 1991) may require a larger value of 
$v_{\rm max}$, provided the plates motion is driven by some other
cause rather than the pull of the fluxoids on the crust.

In contrast to the above {\em velocity} based criterion for
deciding the sense of the bending of the fluxoids, DC93 assume that 
the fluxoids would be bent outward, and become subject to the
same inward force $F_{\rm c}$ as in Eq.~8, throughout the period
of {\em time} \(t < \tau \). They also assume that for times
\hbox{ \( t > \tau \) } an outward force of a comparable
magnitude would be effective. The latter (outward) force will be
however ignored in our new models, since it implies an spontaneous
motion of the end points of a fluxoid at a speed faster than the
fluxoid itself which we don't find it to be justified. The tendency
of a flux tube for decreasing its length under the effect of its
tension (Harvey~et~al. 1986), which could result in an inward
bending, might be of no consequence in the present case. Because,
given the negligible radius of a fluxoid as compared to the radius of 
curvature of the core-crust boundary surface, a fluxoid might not be 
subject to any {\em instantaneous} force pulling it out to a position
with a shorter length; it may remain in its ``meta-stable'' state
being ``unaware'' of the possible gain in its free energy upon an
outward displacement.

On the other hand, the value of $F_{\rm c}$ as given by Eq.~8,
for conditions of \( v_{\rm p} > v_{\rm max} \), might be an
underestimate. It has been argued that the repulsive force
between the fluxoids should ensure that the lattice response to a
deformation is determined, to a first approximation, by their
collective rigidity (Jones 1991).  The force $F_{\rm c}$ associated
with even a piece of the lattice of a size of an inter-vortex
spacing (including some $10^7$ flux lines)
would be, in this approximation, so large that any bending of the
lattice is effectively prohibited.  The velocity of the fluxoids
would be therefore constrained at all times by the condition \(
v_{\rm p} \leq v_{\rm max} \). In order to implement this latter
assumption we, therefore, construct models which use the following
prescription for calculating $F_{\rm c}$. If a value of
\hbox{ \( F_{\rm c} =0 \) } results in deriving \(v_{\rm p}>v_{\rm max}\)
(see below) then \( v_{\rm p} = v_{\rm max} \) is assumed (otherwise,
we set \hbox{\(F_{\rm c}=0\)}, and the derived value for $v_{\rm p}$ 
is used, as before), and $F_{\rm c}$ is calculated from
\begin{eqnarray}  
        F_{\rm c} &=& - \left( F_{\rm n} +F_{\rm b} +F_v \right)
\end{eqnarray}
where the right hand side is evaluated using \( v_{\rm p}= v_{\rm max}
\), together with \( \omega= \omega_{\rm cr} {\ \rm or} \ =
-\omega_{\rm cr}\), whichever may be the case. As indicated earlier,
these latter models might be alternatively viewed as representatives
of the case of coherent electron scattering. 

\section{The Models} 
The steady-state radial motion of a fluxoid, in the region of
interest, is thus determined from the balance equation for all
the radial forces acting on it, per unit length, that is:
\begin{eqnarray}
	F_{\rm n} +F_v +F_{\rm b} +F_{\rm c} &=& 0
\end{eqnarray}
Substituting in the above equation for the different forces from
Eqs~2 or 3, 5, 6, and 8 or 10, respectively, it may be rewritten
in the form
\begin{eqnarray}
 \alpha \ {\omega_{-6} \over P_{\rm s} B_{8}} - \beta \ v_{\rm p_7} + 
\delta = 0 
\end{eqnarray}
where parameters $\alpha$, $\beta$, and $\delta (\equiv F_{\rm b} +
F_{\rm c})$ are given below for the different models, and $v_{\rm
p_7}$ is the fluxoid velocity $v_{\rm p}$ in units of
$10^{-7}$~cm~s$^{-1}$.  Recall that $\omega_{-6}$, which is the value
of $\omega$  in units of $10^{-6} \ {\rm rad \ s}^{-1}$, might have
either positive or negative values, as is also the case with $\delta$
in some of the models.

This single equation includes two unknown variables $\omega$ and
$v_{\rm p}$,
and represents the {\em azimuthal} component of the Magnus
equation of motion (Sonin 1987) for the proton vortices.
No radial Magnus force acts on the fluxoids, hence the right hand
side is set to zero, because of the assumed co-rotation of
the fluxoids with the proton superconductor.  There exist however
additional restrictions on the motion of the fluxoids which can be
used to fix the value of one of the variables and solve Eq.~12 
for the other.  Namely, for a co-moving state \( v_{\rm p} = v_{\rm
n}\) is given and $\omega$ could be determined. In contrast,
when $v_{\rm p}(\neq v_{\rm n})$ is unknown $\omega$ is given as
\( \omega =\omega_{\rm cr} \) or \( \omega = - \omega_{\rm cr} \)
for \( v_{\rm p} < v_{\rm n} \) or \( \ v_{\rm p} > v_{\rm n} \),
respectively. Furthermore, inspection of Eq.~12 indicates that it
admits {\em one and only one} of the three different solutions, 
for the given values of $v_{\rm n}, B_{\rm c}, $ and $P_{\rm s}$
at any time, namely 

\vspace{.5 cm}
\begin{tabular}{llc}
$\omega$ = $\omega(v_{\rm p}=v_{\rm n})$ & \hspace{1.2 cm} iff 
 \hspace{1.2 cm}& $- \omega_{\rm cr} < \omega <\omega_{\rm cr}$      \\
$v_{\rm p}$ = $v_{\rm p}(\omega=\omega_{\rm cr})$ & \hspace{1.2 cm}
iff  \hspace{1.2 cm} & $v_{\rm p} < v_{\rm n}$                       \\
$v_{\rm p}$ = $v_{\rm p}(\omega=-\omega_{\rm cr})$ &  \hspace{1.2 cm}
iff  \hspace{1.2 cm} & $v_{\rm p} > v_{\rm n}$                         \\
\end{tabular}
 
\vspace{.5 cm} 
\noindent
The rate of the
flux expulsion out of the core, $\dot B_{\rm c}= -{2 \over R_{\rm c}}
B_{\rm c} v_{\rm p}$, and the evolution of the stellar surface field
$B_{\rm s}$
(with a decay rate $\dot B_{\rm s} = -{{ B_{\rm s} - B_{\rm c} }
\over \tau}$) are hence uniquely determined from the above
force balance equation, given the spin evolution of the star which
determines (Eq.~1) the vortex velocity $v_{\rm n}$ at any time.

We construct four separate models, labeled A1, A2, B1, and B2, based 
on the two alternative estimates discussed earlier for the pinning
force $F_{\rm n}$ and also for the curvature force $F_{\rm c}$,
by permutation. The models are summarized below by indicating to
which of the four distincting physical assumptions they relate,
and also giving their associated values of the parameters in the
force equation (Eq.~12).
\begin{itemize}
\item Model A1: $\left\{ 
		\begin{array}{ll}
                        & \mbox{{\bf -} vortex {\em segments} creep   
					independently } \\
                        & \mbox{{\bf -} fluxoids may {\em bend} when 
			$v_{\rm p} > v_{\rm max}$ }
		\end{array}  
		\right. $ \\
		$\alpha =  5.03$, \\
		$\delta = \left\{ \begin{array}{ll}
			0.51 & \mbox{ if $ \ v_{\rm p} < v_{\rm max} $} \\
                        0.16  & \mbox{ if $ \ v_{\rm p} \geq v_{\rm max} $}
			\end{array}  \right. $ \\
\item Model A2: $ \left\{ 
		\begin{array}{ll}
			& \mbox{{\bf -} vortices remain {\em straight} while 
                        moving } \\
                        & \mbox{{\bf -} fluxoids may {\em bend} when 
			$v_{\rm p} > v_{\rm max}$ }
		\end{array}  
		\right. $ \\
                $\alpha = 2.587 \times 10^{-4} \ B_8^{1\over2}$, \\
		$\delta =  $ same as for A1
\item Model B1:  $ \left\{ 
		\begin{array}{ll}
                        & \mbox{{\bf -} vortex {\em segments} creep  
					independently  } \\
                        & \mbox{{\bf -} fluxoids remain {\em straight}, 
                        constrained by $v_{\rm p} \leq v_{\rm max}$ }
		\end{array}  
		\right. $ \\
		$\alpha = $ same as for A1, \\
		$\delta = \left \{ \begin{array}{ll}
			0.51 & \mbox{ if $ \ v_{\rm p} < v_{\rm max} $} \\
	- (F_{\rm n} + F_v)  & \mbox{ if $ \ v_{\rm p} = v_{\rm max} $}
			\end{array}  \right. $ \\
\item Model B2:  $\left\{ 
		\begin{array}{ll}
			& \mbox{{\bf -} vortices remain {\em straight}
                        while moving} \\
                        & \mbox{{\bf -} fluxoids remain {\em straight},
                        constrained by $v_{\rm p} \leq v_{\rm max}$ }
		\end{array}  
		\right. $ \\
		$\alpha =  $ same as for A2, \\ 
		$\delta =  $ same as for B1 
\end{itemize}
while $\beta= 7.30$ is the same for all the models \\ 

These four models together with the model adopted by DC93 (the
DCC model) will be referred to collectively as the FBE
models (those which employ a Force Balance Equation), in
contrast to the SIF model which has a different approach for calculating 
the rate of flux expulsion. Table 1 compares all the six models by  
indicating how the forces and the fluxoid velocity are determined in each
of them. Spin and magnetic evolution of the 
single as well as binary pulsars are calculated according to the
requirements of each of the six models, separately, and the results are
discussed in the following sections.

\hspace{-2cm}
\begin{tabular}{|c|c|c|c|cc|c|} 
\multicolumn{7}{l}{\hspace{-.5cm}{ \bf Table 1- Different models
of flux expulsion} }   \\ \hline
\multicolumn{2}{|c|}{model}		& $F_{\rm n}$		& $F_{\rm b}$ and $F_{\rm v}$ & \multicolumn{2}{|c|}{$F_{\rm c}$}	& $v_{\rm p}$ and $\omega$ \\ \hline \hline
		& DCC	&	&    	&as in Eq.~8  \  & if $t < \tau $ &  \\
                &                       &                       &                       & $={\pi \over 4}F_{\rm b}$  \  & if $t \geq \tau$ &    from Eq.~12,    \\ \cline{2-2} \cline{5-6}
	& A1	&as in Eq.~2	&      as in 	 & $=0$  \  & if $v_{\rm p}<v_{\rm max}$ & subject to : \\
                & &     &       &as in Eq.~8  \  & if $v_{\rm p} \geq v_{\rm max}$ &    \\ \cline{2-2} \cline{5-6}
		& B1			&			&  Eqs 	& $=0$  \   & if $v_{\rm p}<v_{\rm max}$  & $v_{\rm p}=v_{\rm n}$ \ iff $ |\omega| < \omega_{\rm cr}$		\\
	FBE	&	&	&	& as in Eq.~10 \ 	 & if $v_{\rm p}=v_{\rm max}$ &   \\ \cline{2-3}  
		& B2			& &5	and 6		& \multicolumn{2}{|c|}{($v_{\rm p} > v_{\rm max}$ not permitted)}  & $\omega = \omega_{\rm cr}$  \ iff $v_{\rm p}<v_{\rm n}$   \\
		&	&as in Eq.~2 \  if $v_{\rm p}=v_{\rm n}$ 	&	& & & \\   \cline{2-2} \cline{5-6}
		& A2			&as in Eq.~3 \  if $v_{\rm p} \neq v_{\rm n}$	&	& $=0$  \  & if $v_{\rm p}<v_{\rm max}$ &	 $\omega = - \omega_{\rm cr}$  \ iff $v_{\rm p}>v_{\rm n}$	\\
                &       & &     &as in Eq.~8  \   & if $v_{\rm p} \geq v_{\rm max}$  & \\ \hline
\multicolumn{2}{|c|}{ SIF}		&\multicolumn{5}{c|}{ 	$v_{\rm p}=v_{\rm n}$ (assumed)}					\\ \hline
\end{tabular}

\vspace{.5 cm}
\section{Single Pulsars} 
The spin evolution of a solitary pulsar driven by its
electromagnetic torque has a rate 
\mbox{$\dot{P}_{\rm s}=3.15\times10^{-32}\
{B_{\rm s}^2 \over P_{\rm s}} \
{\rm s\ yr}^{-1}$}, where $B_{\rm s}$ is the surface field
in units of G, and $P_{\rm s}$ is in units of seconds.
From the instantaneous value of the spin-down
rate one finds the velocity $v_{\rm n}$ of the outward motion of the
neutron vortices (Eq.~1).  Also, the critical lag $\omega_{\rm cr}$
may be determined, from Eq.~4, for any given value of the core field
strength $B_{\rm c}$.  The solution of Eq.~12, for each model,
given the values of $v_{\rm n}$ and $\omega_{\rm cr}$ at a time $t$,  
then determines the corresponding values of the fluxoids outward radial 
velocity $v_{\rm p}$ and the lag $\omega$ between the rotation rates
of the vortices and the neutron superfluid in the core of the
evolving solitary pulsar. The coupled evolution of the spin period and
the magnetic field, in the core and at the surface, are thus
followed over a period of $10^{10}$~yr in order to cover both the
young and the very old neutron stars in the Galaxy.

The computed time evolution of $v_{\rm p}$ and $\omega$ are shown in
Fig.~1, together with $v_{\rm n}$ and $\omega_{\rm cr}$, as is
predicted in the A1 model. Characteristically similar results as
in Fig.~1 are obtained for the other FBE models as well. The evolution
of the lag $\omega$ although is not directly of interest for our present 
analysis of the field evolution nevertheless bears significant 
consequences for an understanding of the rotational dynamics of neutron 
stars, to be discussed elsewhere.

The fluxoids motion in Fig.~1 is seen to follow three evolutionary
phases in which they move slower, together, and faster than the
vortices, successively. These will be referred to as {\em forward}
creep, {\em co-moving}, and {\em reverse} creep phases, respectively
(we use the terminology of DC93). Transitions between these
successive evolutionary phases occur because of the reduction in
$v_{\rm n}$ ($\propto \dot \Omega_{\rm s}$) as well as the increase
in $P_{\rm s}$, and a final co-moving phase might also occur for
some choices of the initial conditions. Note that $\omega$ changes
sign from positive to negative and remains so in the later parts
of the co-moving phase, as well as during the reverse creep phase.
Also note that $|\omega| = \omega_{\rm cr}$ during both the
forward and the reverse creeping phases.

\subsection{Field Evolution}
The predicted evolution of the core and surface fields for a single
neutron star according to the A1 model is given in Fig.~2; the other
FBE models produce similar results. The two panels in Fig.~2
are for two different assumed values of the decay time scale
$\tau$ in the crust, where Fig.~2a corresponds to the results in Fig.~1. 
A substantial decrease in the core field occurs at a time $ t \gtrsim
10^7$~yr, which is expected for the typical average values of $v_{\rm
p} \lesssim 10^{-8}\ {\rm cm \ s}^{-1}$ during the earlier times,
because \( \frac{\dot B_{\rm c}}{B_{\rm c}} =
\frac{v_{\rm p}}{R_{\rm c}} \) implies that a time period \( \Delta
t \sim \frac{R_{\rm c}}{v_{\rm p}} \) is needed for a major reduction
in the core field to occur. However, because of the very small
magnitude of $v_{\rm p}$ (although $\gtrsim v_{\rm n}$) and also
the reduced value of $B_{\rm c}$ at later times $B_{\rm c}$ does
not change, substantially, afterwards.
The surface field $B_{\rm s}$ responds to the change in $B_{\rm c}$ on
the assumed decay time scale $\tau$ of the crust. The nontrivial role 
of the stellar crust in these field evolution models may be seen by
comparing Fig.~2a with Fig.~2b, where values of $\tau = 10^7$,
and $10^8$~yr have been used, respectively. A larger value of
$\tau$ tends to maintain the initial $B_{\rm s}$, hence a larger 
$\dot P_{\rm s}$ as well as a larger $v_{\rm n}$, over a more extended
period of time.  Consequently, smaller final values of $B_{\rm c}$
and $B_{\rm s}$ are predicted for the larger assumed values of
$\tau$, as is seen in Fig.~2. 

Fig.~2a might seem to suggest that $B_{\rm c}$ stops decaying
once $B_{\rm s}$ starts to decline, as is also concluded by DC93.
However, this is but an artifact of the assumed value of $\tau$
which happens to be close to the saturation time of the core field
decay. The spurious nature of such a correlation, that a
decrease in the strength of the crustal field results in 
a halt in the decay of the core field, may be confirmed by
using tentatively smaller values of $\tau \lesssim 10^6$~yr.
As we have verified, in such cases the core
and the surface fields are seen to decay simultaneously
(see Fig.~3.3 in Jahan-Miri 1996b), in contrast to the above
unreal correlation. The decay of the core field, according
to the present models, does not have a direct and one-to-one
dependence on that of the crust; instead they are coupled
through the influence of many factors (see the
preceeding paragraph). This statement may be also appreciated
by contrasting the predicted field evolution under distinct 
intial conditions for the relative fields of the core and that
of the crust, $B_{\rm crust}$. For this purpose, 
the results in Fig.~2 with an assumed intial condition
$B_{\rm crust} \sim 0.1 B_{\rm c}$ may be compared with that
in DC93 where $B_{\rm crust} = 1.5 B_{\rm c}$ and also
$B_{\rm crust} \sim 5 \times 10^3 B_{\rm c}$ have been used
(their Figs~2 and 3). The initial crustal field in our case
compares with a very late stage of evolution in DC93 when
the crustal field has been almost completely decayed. Hence, 
the initial core field in Fig.~2 should have not decayed, from
the very beginning, if the above effect were real.
The above differecnce in the adopted initial fields might
deserve a further note. For an initially uniform
distribution of the magnetic flux within the star, which is
assumed in DC93 as well as here, the initial condition
$B_{\rm crust} = 1.5 B_{\rm c}$ implies a relative radial
size for the crust larger than the currently used values
(eg., Sauls 1989; Pines \& Alpar 1992). It is particularly
questionable that the same condition has been used 
in DC93 for their various choices of EOS that have
different predictions for the relative size of the crust.

\subsection{Force Analysis}
The time evolution of the radial forces acting on the fluxoids
is shown in Fig.~3, for the A1 model (corresponding to the results
in Fig.~1
and Fig.~2a). One should note that a fluxoid, being a vortex,
responds to a force by moving in a direction perpendicular to the
force; a radial force does {\em not} affect its radial motion,
directly. However, because of the dependence on $v_{\rm p}$ in Eq.~12 
(which originally describes the azimuthal motion of the fluxoids)
a ``driving'', or a ``braking'' role in flux expulsion might be
assigned to an outward directed (positive), or an inward directed
(negative) force, respectively.

As is seen in Fig.~3, the pinning force $F_{\rm n}$ is negative
(directed inward) during the reverse creep phase and also the
later part of the co-moving phase. Nevertheless, the major
predicted flux expulsion
does occur (see Fig.~1 and Fig.~2a) during the co-moving and,
particularly, the reverse creep phases. This means that the dominant
``driving'' force for the flux expulsion is the buoyancy force
which is positive throughout the evolution. And, that the overall
role of the pinning force in 
the field decay of solitary pulsars is more like a ``brake'' {\em
preventing the flux to be expelled too rapidly}. We have further
verified this conclusion, which seems to be obvious from the
results in Figs~1--3, by other tests of the model calculations where
we have tentatively set either of the two forces equal to zero.
Our conclusion about the braking role of the pinning force is,
however, in contradiction with that of DC93 who attribute a
driving role to this force throughout their paper, and make
further statements which are not hence justified. The braking 
role of the pinning force is also
in sharp contrast with the view adopted by Srinivasan et~al.
(1990) about the flux expulsion mechanism, eventhough comparison
with their (SIF) model, in this regards, might be misleading
and unjustified since their treatment does not address the
dynamics to begin with.

The opposing role of the pinning force against the fluxoids
outward migration, at late times ($\gtrsim 10^7$~yr), is indeed
true for all the FBE models adopted here. Except for
including the pinning force these models are, otherwise, similar
to the earlier dynamical studies of flux expulsion
(Muslimove \& Tsygan 1985; Harvey~et~al. 1986) in having the
buoyancy force as the driving cause of the flux expulsion, as
verified above. It is for the role of the pinning force that 
the predicted field of a neutron star, for all of the FBE models, 
may never decay to very small values ($< 10^8$~G) even after very
long times ($\gtrsim10^{10}$~yr). This is a fundamental difference
which is much needed to account for the low field pulsars (to be
discussed in the next section) that could not be possibly explained
by the earlier flux expulsion scenarios which neglected the pinning.

We hoped to be able to comment about the relevance of the
alternative
physical assumptions discussed for the FBE models by comparing
their
predictions. However the predicted field evolution behaves very
similarly in the different models, hence no further conclusions
may be drawn. The reason for this rather unfortunate finding
may be retraced to the braking role of the pinning
force, $F_{\rm n}$, as well as the effective period of the core
field expulsion; both discussed earlier. Models A1, B1, and DCC differ from
the other two (see \S{3}) in their prescription for calculating
the
magnitude of $F_{\rm n}$. Due to the dependence of $F_{\rm n}$
on $B_{\rm c}$ and $P_{\rm s}$ its predicted value for the
two classes of models is markedly different only during the
early short phase of forward creeping (see, also, Fig.~3.4 in
Jahan-Miri 1996b). However, the major field expulsion occurs
during the later prolonged phases, ie. when $F_{\rm n}$ acts as
a brake and its predicted value is not much different for
the different models. Hence, the distinction among the models
with respect to $F_{\rm n}$ is, in effecte, washed out. 
A second distinction among the models
(models A versus B) relates to their different prescriptions for
calculating the curvature force, $F_{\rm c}$, which differ
only when the fluxoid velocity is large ($\gtrsim v_{\rm max}$). 
Large fluxoid velocities, which occurs only when neutron
vortices too move fast, hence large differences between the
value of $F_{\rm c}$ among the models takes place again only
during the early phase of rapid spinning down of a single
pulsar. Thus, in short, for the spin-down history of single
pulsars the calculated forces on fluxoids at times $>1$~Myr 
are not much different for the different models treated here,
hence resulting in similar field evolutions. 
The distinction among the predicted field evolutions due to
the different FBE models could however be, in principle, quite
significant as is encountered in the case of spin histories
of neutron stars in close binaries which is discussed below. 

\subsection{Observational Implications}

\vspace{.5cm}
\noindent
{\bf Pulsar distribution:} \\
The evolutionary tracks for single pulsars on the spin-magnetic
field diagram as predicted by SIF and DCC (the latter being
typical for all the FBE models) are plotted in Fig.~4, for the
different assumed initial field strengths. Points corresponding
to the given ages of the neutron star are also marked along
each track.  As is seen in Fig.~4, the predicted final strength
of $B_{\rm s}$, for FBE models, is found to depend sensitively,
and inversely, on its initial value. This is a consequence of
the direct correlation between the total expelled flux and the
initial value of $B_{\rm s}$, which is expected because larger
values of $P_{\rm s}$ are achieved for larger initial $B_{\rm s}$
values, as indicated earlier. However the final value of
$B_{\rm s}$ is insensitive (not shown in Fig.~4) to the assumed
initial values of $P_{\rm s}$ and $B_{\rm c}$, for changes in
these quantities by almost two orders of magnitudes.
In contrast, for the SIF
model the final value of $B_{\rm s}$ is found to
depend on the initial values of $P_{\rm s}$ and $B_{\rm c}$,
as well as on the initial $B_{\rm s}$ (Fig.~4) eventhough having
a direct
correlation in this case. These correlations, for the SIF
model, are in accord with the assumed relation \(\frac{\dot
B_{\rm c}}{B_{\rm c}} = - \frac{\dot P_{\rm s}}{P_{\rm s}}\),
corresponding to $v_{\rm p} = v_{\rm n}$ at all times.
A further point to note in Fig.~4, is the power-law  time
behavior of the field evolution of single pulsars at late times,
which is realised for the lower initial fields. This feature
which is common among the SIF and the FBE models, and was
also pointed out by Srinivasan et~al. (1990), has new
observational implications which have been previously
highlighted (Jahan-Miri 1996a).

\vspace{.5cm}
\noindent
{\bf Very old neutron stars :} \\
In spite of the recent discovery of large redshifts for some 
$\gamma$-ray burst sources the association of a sub-class of
them with a galactic population of highly magnetic old neutron
stars (see, eg., Blandford 1992) may still be viable. While such an
identification of the $\gamma$-ray bursters does not seem to be
consistent with the predictions of FBEs for the field strengths
of very old single neutron stars, it could be however accommodated
by the SIF model.
According to SIF very old neutron stars (with ages $\sim 10^{10}$)
are expected to have rather large magnetic fields in the range
\(2 \times 10^{10} \lesssim B_{\rm s} < 2 \times 10^{11} \)~G, while the
FBE models predict values of \( B_{\rm s} \lesssim 2 \times 10^{10} \)~G
for such
stars (see Fig.~4). The FBE-predicted upper limit for the final fields
is even smaller than above, $B_{\rm s} < 3 \times 10^9$~G at an age
$ \gtrsim 10^8$~yr, for the large initial values of
$B_{\rm s} \gtrsim 10^{12.5}$~G. Furthermore, while the above results from
Fig.~4 are for a value of  $\tau =10^7$~yr, still smaller final fields
would be the case for larger values of $\tau$ (compare Fig.~2a with
Fig.~2b).

In contrast, DC93 suggested that for a neutron star having a 
magnetic axis aligned with its rotation axis the core field,
according to the DCC model, would not be
expelled even on large time scales ($>> 10^7$~yr), unlike the general
case presented in Fig.~4. Thus, they concluded that the earlier proposed 
model of Ruderman \& Cheng (1988) for the burst sources being
aligned neutron stars is consistent with a field evolution
according to their (DCC) model. However, their conclusion has
to be dismissed since it is based on the unusual assumption that
the spin-down torque of a pulsar is only due to its magnetic
dipole radiation. The usual and long lived consensus which is
commonly adopted also for an observational determination of the
strengths of the surface fields of pulsars is that the spin-down
torque, due to combined effects of dipole radiation and outflow of
relativistic particles, is independent of the inclination angle
(see, eg., Manchester \& Taylor 1977, pp. 176-180; Srinivasan 1989).

Nevertheless,
since a discussion of the predicted field decay by the FBE models for
a {\em tentatively} assumed case of little spin-down torque acting on
a neutron star serves to further elucidate the nature of the models we
will pursue the discussion. Indeed, if the spin-down rate of a neutron
star is assumed to be small the core field according to the FBE models
would not be expelled much, as suggested by DC93. This happens mainly 
because a reverse creep phase does not occur in this case since the
associated value of $|F_{\rm n}(\omega= - \omega_{\rm cr})|$ 
would be too large. The large value of $F_{\rm n}$ follows from
its inverse proportionality on $P_{\rm s}$ (see Eq.~2), and the
fact that the assumed small spin-down torque result in small final
values of $P_{\rm s}$. However, Ding~et~al. argued (see the last equation
in DC93) that the large value of $F_{\rm n}$ in this case is a
consequence of the direct proportionality of $\omega_{\rm cr}$ on
$B_{\rm c}$, and the large value of $B_{\rm c}$. We argue that the
dependence of $\omega_{\rm cr}$ on $B_{\rm c}$ is irrelevant since 
$|F_{\rm n}(\omega= - \omega_{\rm cr})|
\propto {1}/{B_{\rm c}^{1\over2}}$, which implies a smaller 
$|F_{\rm n}|$ for the assumed larger $B_{\rm c}$.  In fact the
reverse creep phase starts always at a large value of $B_{\rm c}$ even
in (the non-aligned) cases where a substantial field decay does occur,
as might be expected from the above dependence of $F_{\rm n}$ on
$B_{\rm c}$. We conclude that the FBE-predicted little flux expulsion
for the ``aligned'' case is only because $P_{\rm s}$ could retain its assumed
small initial value, as the following test verifies. 
The two cases of field evolution presented in Fig.~5 are both for an
assumed very small spin-down torque (ie. that expected due only to the 
dipole radiation for an inclination angle of $1$~deg) but for two 
different initial values of $P_{\rm s}$.  Eventhough in both
cases $B_{\rm c}$ is large before the transition to the reverse
creep phase, however substantial flux expulsion does takes place
in the case where $P_{\rm s}$ is
large.  Note that in this latter case the reverse creep phase starts
from the very beginning and persists throughout the evolution of the
star.  This clearly verifies the above mentioned role of $P_{\rm s}$
in determining the conditions for occurrence of a reverse creep phase.
In addition, the above arguments indicate a potential possibility for
having very old neutron stars with strong magnetic fields, in the
context of the FBE models too. This may be achieved provided the star is 
not spun down to large periods $ \gtrsim 1$~s, independent of whether
or not the surface field is aligned with the rotation axis of the
star, as it also happens to some extent for the low field pulsars
shown in Fig.~4.

\vspace{.5cm}
\noindent
{\bf Active lifetimes of pulsars :} \\
The radio-active lifetimes (defined by the condition
$\frac{B_{\rm s}}{P_{\rm s}^2} = 0.2 \times 10^{12}$) of pulsars,
calculated for the predicted spin-field evolution in the different
FBE models, are shown in Fig.~6 against the initial field strengths,
for the two values of $\tau =10^7 \ {\rm and } \  10^8$~yr, separately.
Corresponding curves for the case of exponential field decay model,
labeled as ``Exp.'', 
(which assumes the total surface field decays exponentially, with
no constraint, on the given time scale $\tau$) are also included, for
comparison. The predicted lifetimes by the different FBE models are
very similar as can be seen in Fig.~6 (this is true also  for the
A2 and B2 models which have been omitted for the clarity), again
making any attempt for a selection among the five FBE models
fruitless. Nevertheless, there exist a marked difference (in Fig.~6)
between the results of FBEs, as well as SIF, in contrast to those
of ``Exp.'', in particular for values of $\tau \lesssim
10^7$~yr. This difference should, in principle, have testable
consequences for studies of pulsar statistics to decide between
the flux expulsion scenario, in general, and the exponential model,
for the single pulsars (see Jahan-Miri, 1996a, for further discussion).

\section{Binary Evolution Models}
The spin evolution of a neutron star in a binary system with a
mainsequence star is expected to be different from that of
a single pulsar. The interaction of the neutron star magnetosphere
with the stellar wind of the companion star could result in final
large values of $P_{\rm s} \sim 10^4 $ -- $10^5$~s, in contrast to
the much smaller values achieved in the case of single
pulsars. Magnetic evolution of binary neutron stars as predicted
by the flux expulsion models (FBEs and SIF) is therefore
expected to be, in principle, quite different than that of
the single pulsars. In this section we employ the FBE models, for
the first time, in a study of the field evolution of neutron
stars in binaries. We consider models for the orbital and spin
evolution of a neutron star, of a mass $M_{\rm n}$, born in a
binary with an orbital period $P_{\rm orb}$, corresponding to
an orbital separation $a$. The binary companion of the neutron
star is a mainsequence star of mass
$M_2$ which loses mass in the form of a
spherical uniform stellar wind at a rate $\dot {M}_2$.
In the picture of flux expulsion models the evolutions of the
spin period
and the magnetic field of the neutron star in such a binary would be
intimately coupled.  While the spin-down process would tend to
reduce
the field strength, the reduced field strength (together with the
increased spin period) will in turn affect the rate and the direction
of the spin variations. We follow this coupled evolution of the
surface magnetic field and the spin period of the neutron star for a
time of the oreder of the expected mainsequence lifetime of
the companion
star. The orbital and spin evolutions for the expected
Roche-lobe overflow phase is not however simulated.
Nevertheless, it should be noted that the computed field
evolution of the
recycled pulsars is not affected by this omission of the
Roche-lobe phase of the binary evolution. The latter
spin-up phase of a recycled pulsar would have no direct effect
on the flux expulsion, hence no consequences for the field
evolution, except for the general decay of the crustal field
which is accounted for in our computations. Thus our simulations
are complete, within the limitations of the models, as far as
the field evolution of recycled pulsars is concerned; a point
which has been overlooked by some authors (Urpin et.~al. 1998)
in referring to the earlier cited binary evolution simulations
based on the SIF model.

We assume that the first phase of the binary evolution,
namely the active-pulsar phase with a dipole spin-down
(during which the neutron star spins down due only to the
dipolar radiation torque on it), lasts till the ram pressure
of the stellar wind overcomes the pressure of the ``pulsar
wind'' at the accretion radius
(see, e.g., Illarionov \&
Sunyaev 1975; Davies \& Pringle 1981).
During this period the stellar wind will have no dynamical
effect on the neutron star.  The pulsar's core
magnetic field will undergo an expulsion determined by the
dipole spin-down rate, according to the FBE models.  In the
subsequent two phases where the accreted wind
matter interacts directly with the magnetosphere we assume
that a steady Keplerian disk is formed by the accretion flow
outside the magnetosphere, with the same sense of rotation
as that of the neutron star.  This is the least efficient
configuration for angular momentum extraction from the
neutron star which we have considered, in addition to the
more efficient geometries such as a spherically symmetric
radial infall.

The accretion flow interacts with the magnetosphere at its 
characteristic boundary radius $R_{\rm mag}$, which is
defined by the condition of balance between the magnetic
pressure and the ram pressure of infalling flow (Davidson
\& Ostriker 1973):
\begin{eqnarray}
 R_{\rm mag} &=& 1.88 \times10^{-12}\;\;\; \left(
\frac{B_{\rm s}^2}{\dot{M}_{\rm acc}}\right)^{2/7} \ \ R_{\odot}
\end{eqnarray}
where $\dot M_{\rm acc}$ is the rate of capture of wind matter,
in units of solar masses per year, as defined in Jahan-Miri \&
Bhattacharya (1994; Eq.~4 therein).  This interaction spins
the neutron star up or down depending on the sign of the quantity
$V_{\rm dif}(=V_{\rm corot}-V_{\rm Kep})$ evaluated at the
boundary of the magnetosphere.  Here, $V_{\rm corot}$ is the speed of co-rotation
with the neutron star at a given distance from it, and $V_{\rm Kep}$ is
the Keplerian speed at the same distance.
In the limiting case when the
co-rotation velocity $V_{\rm corot}$ becomes equal to the Keplerian
velocity $V_{\rm Kep}$ the neutron star will conserve its spin period
while accretion onto the star will continue.  The rate
$\dot L_{\rm s}$ of transfer of angular momentum between the
stellar wind and the neutron star
is assumed to be equal to $\dot{M}_{\rm acc}$ times a
specific angular momentum corresponding to the difference between the
co-rotation velocity with the neutron star and the Keplerian velocity 
evaluated at a distance $R_{\rm mag}$ from the neutron star.
In general, then, 
\begin{eqnarray}
 \dot L_{\rm s}=\eta \times V_{\rm dif} \times R_{\rm mag} \times 
\dot{M}_{\rm acc} 
\end{eqnarray}
will be used, where $\eta$ is the efficiency factor included to take
into account the uncertainties due to the detailed geometry of the
interaction and the actual value of the specific angular momentum
carried by the accreted wind just before and after the interaction.
Different rates of angular momentum transfer are thus tested by
assuming different values for $\eta$ while $\eta=1$ corresponds to the
case of disk-accretion with a mechanical torque acting at
$R_{\rm mag}$. Jahan-Miri (1996a) found that in order to account
for the properties of binary pulsars evolved in both massive as well
as low-mass systems, simultaneously and self-consistently, larger
values of $\eta >1$ were preferred. Hence, and in contrast to the
adopted range of values between $0.2$--$1.0$ for the efficiency
factor in Jahan-Miri \& Bhattacharya (1994), here we use the larger 
values of $\eta$; see below. 

The following coupled differential equations for the time
evolution of $M_{\rm n}$, $a$, $P_{\rm s}$, along with those
for $B_{\rm c}$ and $B_{\rm s}$ discussed before (\S{3}), 
are solved numerically, for the various combinations of
parameter values indicated below.
\begin{eqnarray}
\frac{{\rm d}a}{{\rm d}t} &=& 2a\left\{\frac{\dot L_{\rm losses}}{L_{\rm orb}}\Bigm|_{\rm losses}- 
\frac{\dot{M_2}}{M_2}\left[1+(\alpha-1)\frac{M_2}{M_{\rm n}}-\frac{1}{2}
 \alpha \frac{M_2}{M} -\alpha\beta\frac{M_{\rm n}}{M} \right]\right \} \\ 
 		& \ & \nonumber \\
 \frac{{\rm d}M_{\rm n}}{{\rm d}t} &=&\cases {\dot{M}_{\rm acc} \hskip 1cm 
        \mbox{$\ldots  \ldots \ldots  \ldots$ 
        \hskip 0.5cm \rm accretion phase}\cr
        0.0  \hskip 1.2cm   \mbox{ $\ldots \ldots \ldots \ldots 
       $  \hskip 0.5cm   \rm propeller phase}\cr} \\
       		& \ & \nonumber \\
\frac{{\rm d}P_{\rm s}}{{\rm d}t} &=& 3.18\times10^{-3}  \; \eta  
 \left(\frac{\dot{M}_{\rm acc}}{M_\odot{\rm yr}^{-1}}
 \right)\left({R_{\rm mag}\over{\rm km}}
 \right)\left(\frac{P_{\rm s}}{\rm s}\right)^2\left(
 \frac{V_{\rm dif}}{{\rm km \ s}^{-1}} \right) \
 \mbox{\rm s yr}^{-1}
\end{eqnarray}
where $L_{\rm orb}$ is the orbital angular momentum,
$\dot L_{\rm losses}$ is the rate of change in $L_{\rm orb}$
except for the contribution due to the escaping matter from
the system which is already taken into account,
$M= M_{\rm n} + M_2$ is the total mass of the binary,
$\alpha$ is the ratio of the mass loss rate from the system to that
from the secondary, and  $\beta$ is the ratio of the effective specific
angular momentum of the escaping matter to that
in the companion star. As indicated earlier, a spin-up phase
of the neutron star would have no effect on the flux expulsion
out of its core, and the core field remains constant during
such a period of time. Also, note that during the active
pulsar phase $\dot P_{\rm s}$ will be
given by the relation indiacted earlier (\S 4) for the single
pulsars, instead of Eq.~17 which is applicable for the other 
two phases of magnetospheric interaction with the accreted
matter; see Jahan-Miri \& Bhattacharya (1994), and Jahan-Miri
(1996a), for more details of the binary evolution models. The
computations were repeated using different combinations of the
following values of the parameters and the initial conditions,
for each of the FBE models, separately:

\bigskip
\vbox{\halign{\quad #\hfil &\quad #\hfil \cr
$P_{\rm orb}$: 2 -- 600                  &(day) \cr
$\eta$: 1, 10, 100                        &{ } \cr
$\log \tau$: 7.0, 8.0, 9.0               &(yr) \cr
$\log \dot M_2$: $-$15, $-$14, $-$13       &($M_{\sun}$~yr$^{-1}$) \cr
initial $P_{\rm s}$: 0.1, 1.0 &(s) \cr
initial $B_{\rm s}$:  $3\times10^{12}$   &(G) \cr
initial $B_{\rm c}$:  $2.7\times10^{12}$ &(G) \cr}}
\bigskip

\noindent
The evolution of a neutron star in a binary with a
low-mass companion ($M_2=1.0 \ M_{\sun} $) is followed for a
period of $10^{10}$~yr and its final surface field is
determined, in order to be compared with the observed fields
of low mass binary, and millisecond, pulsars.

\section{Predicted Fields of Recycled Pulsars}
The general features of the computed evolution for the outward
velocities of the fluxoids and the vortices, as well as the
rotational lag, are similar to that descibed earlier in the
case of single pulsars. A new feature is the rapid increase in
the fluxoids velocity during the reverse creep phase, which
is expected because of the enhanced spinning down of the star 
in a close binary. 
The distribution of the final surface field strengths versus the
initial orbital periods are plotted in Fig.~7, as predicted by two
of the FBE models, for the given parameter values.
Fig.~7 shows that the observed magnetic field strengths of the
low-mass recycled pulsars, and millisecond, pulsars may be 
successfully reproduced by the FBE models.
In addition, the observational
data on eight low-mass binary pulsars, which are expected to have been
recycled in low-mass binary systems, are also presented in Fig.~7.
The measured orbital periods for these systems have been corrected
for the expected
change in the period during a final Roche-lobe overflow mass transfer
phase in the binary to infer the corresponding intial values
of $P_{\rm orb}$, listed in Table~1, for use in Fig.~7.

\vskip 1cm 
\begin{tabular}{cccccc}
\multicolumn{5}{l}{\hspace{-.5cm} {\bf Table 1- The observed low-mass binary pulsars} } \\ \hline
PSR& $P_{\rm s}$& $P_{\rm orb}$& $M_2$    & $\log B_{\rm s}$& initial-$P_{\rm orb}$ \\
   & (msec)     & (day)        & $(M_\odot)$&      (G) &  (day) \\ 
\hline
0820+02&     864&          1232&   0.2--0.4&     11.48&   300 \\
1953+29&     6.13&           117&   0.2--0.4&      8.63&    12 \\
1855+09&     5.36&          12.3&   0.2--0.4&      8.48&     1 \\
J1713+0747&  4.57&         67.83&   0.3--0.5&      8.28&     6 \\ 
J2019+2425&  3.93&         76.51&   0.3--0.5&      8.26&     7 \\ 
J1643$-$1224&  4.6&          147&        0.14&       8.6&     24 \\
J1455$-$3330&  8.0&           76&         0.3&       8.3&     10 \\ 
B1800$-$27&    334.4&        406.8&       0.15&      10.9&    60  \\
\hline
\multicolumn{1}{r}{\small References:}&
\multicolumn{5}{l}{\small Bhattacharya and van den Heuvel 1991;} 
{\small Foster, Wolszczan and Camilo 1993}; \\
 & \multicolumn{5}{l}
{\small Nice, Taylor and Fruchter 1993;}
 {\small Johnston~et~al. 1995; Lorimer~et~al. 1995}.\\
\end{tabular}

Within the uncertainties associated with the value of
$\dot M_2$, which could be also varying with time,
and the other unknown parameters of the binary pulsars, 
the computed curves in Fig.~7 seem
to agree with the data points. It is also understood that while
the curves in Fig.~7 represent a particular choice of the values
for the pulsar-binary parameters however the true value of each
of them might have been quite different among the corresponding
eight systems. Qualitatively similar agreement with the data, as
in Fig.~7, is obtained also for many other choices of the
parameter values, as well as for the other FBE models, namely
A2, B1, and B2. Nevertheless, and in contrast to the case of
single pulsars, the predicted field evolution of a given
binary pulsar is found to be quite different according to 
the different FBE models. This is promissing, in the sense 
that it offers a potential possibility not accessible from the
application of the models to single pulsars. One
might hope to distinguish among the models and
gain insight into the interior physics of neutron stars, based
on a comparison of the predicted spin-field evolution with
the observational data on the recycled systems.
However, we have not been able to pinpoint any preferences 
among the various flux expulsion models, at this stage,
because of the uncertainities due to the free and unknown
parameters (ie. $\eta$, $\dot M_2$, $\tau$, etc.) which are
encountered even in a simplified treatment of the binary
evolution that we have used.
That is, within the indicated ranges for the parameter values,
all FBE models merit the same success in comparison to the 
existing observational data on the recycled pulsars.

Nevertheless, the above general success of FBE, and
SIF, models which has not been so far reported for any other
field decay model of neutron stars, provides a strong support
for the flux expulsion scenario. It is further noted that
the SIF
model was shown previously to be consistent with the data  
also in the case of pulsars recycled in binaries with
{\em massive} companion stars (Jahan-Miri 1996a). 
Thus, judging on the overall agreement seen here between 
the predictions of SIF with that of FBE models,
in the case of single and low-mass binary pulsars, it
might be justified to generalize the success of SIF for 
the case of massive binaries to FBE models as well.
On the other hand, the other
existing field evolution model of neutron stars, which
assumes the total magnetic flux to be confined to the crust,
has been also shown to result in the case of binary pulsars
to a decay of the field down to values similar to that of
the recycled pulsar (Geppert et.~al. 1996; Urpin et.~al. 1998). 
However, the spin-orbital binary evolution have not been
really followed in these studies in all its details, and
instead general
estimated values for the rate and duration of accretion
of matter onto the neutron star have been used. More
importantly, the dependence of the final fields on the
orbital periods which is shown in Fig.~7 for the FBE models
and is also compared with the observational data remains to be
demonstrated for the model of crustal field decay due to
accretion.

As in the case of single pulsars, for neutron stars evolved in
many of the binary systems which we have simulated, again the
pinning force on fluxoids acts as an obstacle against an otherwise
more rapid and enhanced flux expulsion. Indeed, we have verified
that setting $F_{\rm b} + F_{\rm c}=0$ in the models, namely having
$F_{\rm n}$
as the only existing driving force, results in a much smaller flux
expulsion than otherwise. In contrast, if $F_{\rm n}=0$ is adopted 
practically zero final field values are obtained, which further demonstrate
the braking role of $F_{\rm n}$. Nevertheless, the essential role
played by the pinning force in the field evolution of recycled old
pulsars may not be overlooked. A model which discards the
pinning force and relies only on the buoyancy is obviously unable
to account for any flux to be present in the cores of such old 
pulsars. In contrast, the FBE (and SIF) models not only 
account for the observed fields of the old binary and millisecond
pulsars, they also predict a correlation between the final field
strength of a recycled pulsar and its spin period history, by virtue
of the role of the pinning force.

\section{Summary and Conclusions}
We have studied a scenario for the evolution of the magnetic
fields of neutron stars, assuming that the magnetic flux resides
in the proton superconductor core of the star and is carried by
the fluxoids. We have employed a dynamical treatment for 
the flux expulsion trying to improve our earlier studies
based on the original model which adopted 
an expulsion rate equal to the spin-down rate of
the star. In the present
work, the rate of expulsion of the flux out of the core is
determined by explicitly calculating the radial velocity of
the fluxoids which are subject to various forces.  We have included
forces due to pinning between the fluxoids and the neutron superfluid 
vortices, buoyancy, scattering of electrons, and tension of the flux
lines. Alternative possibilities for evaluation of these forces
were considered. The predictions of the various corresponding
field decay models for the evolution of single and binary pulsars
were discussed, and our conclusions are further summarized below.

\begin{itemize}
\item 
The explored flux expulsion models predict a restricted decay of
the magnetic field of pulsars that also accounts for the residual
fields of the very old neutron stars. The pinning force plays two
opposite and essential roles in the magnetic evolution of the star.
The success of the models in predicting the residual fields of the
recycled binary and millisecond pulsars is because of the braking
role of this force against the fluxoids outward motion. This effect
also result in a long-lived slowly evolving phase, hence an
increased radio-active lifetime, for some of the single pulsars.
However, the pinning force also has a positive role in driving the
fluxoids out of the core which is revealed in the predicted
dependence of the final field strengths of the recycled pulsars
on the binary parameters that determine the spin history of the
star.
\item
Flux expulsion under the influence of the buoyancy force alone,
in the absence of the pinning force, leads to vanishing field
strengths in old binary and millisecond pulsars, hence it is
definitely ruled out. In contrast, a (tentative) model which
invokes the pinning force and neglects the buoyancy would
result, in general, in larger final fields than in the presence
of the both forces; however, for some choices of the parameters
the final fields might be small as well. 
\item 
The alternative dynamical models, which we have considered
for the motion of fluxoids and vortices in the core of neutron
stars, result in similar predicted spin-magnetic evolutions for 
the single pulsars. For the binary pulsars,
eventhough the predictions of the models are quite different
in many of the cases, however similar acceptable results are
still obtained for each of them, albeit for the different
plausible values of the parameters. Hence, the underlying
assumptions in these models remain observationally equivalent.
The ideas which we entertained in these models are {\em a)} a
neutron vortex
remains straight while moving, versus, its pinned segments might
creep independently, {\em b)} a fluxoid may bend and be reacted
by its tension, versus, the collective rigidity of the lattice
prevents any bending, {\em c)} fluxoids are bent outward (inward)
at all times smaller (larger) than the field decay time scale in
the crust, versus, fluxoids are bent outward when moving faster 
than the permitted motion of their end points due to diffusion of
the flux in the crust and remain straight otherwise. 
\item
The role of the coherent scattering of the electrons by the fluxoid
lattice does not seem to be fully understood. Flux expulsion in
the presence of the coherent scattering might depend on the electron
currents across the core-crust boundary. Assuming that
the associated maximum radial velocity of the electrons is similar
to the limiting velocity of the fluxoids in the case of collective
rigidity of their lattice, we speculated that our models for the
case of collective rigidity would represent the case of coherent
scattering too.
\item 
In contrast to the dynamical models which we used to determine the
time evolution of the fluxoids velocity, the original spin-down
induced flux expulsion model assumes equal velocities for the
fluxoids and the vortices, at all times. The close agreement between
its predictions and those of the dynamical models
is largely because a substantial flux expulsion occurs during a
co-moving phase of the two families of the vortices, even according 
to the latter models. This model, which offers a much simpler
approach for a calculation of the field evolution compared to that
employed here, might be further justified on the dynamical grounds. 
The co-moving phase corresponds to a less energy dissipation
for the coupled system of the two families of vortices, than their
crossing through each other which costs energy. It is feasible,
though need to be demonstrated, that the more economical co-moving
phase is preferred, and maintained at all times, by the
magnetohydrodynamics of the interior quantum liquid; noticing also 
that the fluxoids as well as the vortices are both magnetized and
are rooted in the same highly conductive medium at the bottom of
the crust.
\item 
Finally, an effective time scale in the range $10^7$--$10^8$~yr for the
decay of the magnetic field in the crust of a neutron star is
suggested, based on the preferred results of the model calculations
presented here.
\end{itemize}

\acknowledgments

The author is grateful to the referee for the very useful
comments and suggestions which helped to improve the text largely. 
I wish to thank G. Srinivasan for reading a separate earlier
report of this work and making useful suggestions. I am in debt to
D. Bhattacharya for use of the binary evolution code which was 
developed for our earlier work. I am grateful to
Raman Research Institute for their kind hospitality, and for
using their Computing facilities in carrying out this study.

\clearpage

\figcaption{The {\em top} panel shows the predicted time
	evolution of the lag $\omega$ and its critical value
	$\omega_{\rm cr}$ in a solitary neutron star according to the
	A1 model. The {\em bottom} panel shows the corresponding
        evolution of the velocities of the fluxoids ${\rm V}_{\rm p}$,
        and the vortices ${\rm V}_{\rm n}$.  Initial values of $B_{\rm
	s}=10^{12.5}$~G, $B_{\rm c} = 0.9 B_{\rm s}$, and a value of
        $\tau = 10^7$~yr have been used.}

\figcaption{The predicted time evolution of the strength of the
	magnetic field in the core $B_{\rm c}$ and at the surface
	$B_{\rm s}$, and the spin period $P_{\rm s}$ in a solitary
        neutron star according to the A1 model. The {\em top} panel
        {\bf (a)} is for an assumed value of $\tau = 10^7$~yr and
        corresponds to the results in Fig.~1, while the {\em bottom}
        panel {\bf (b)} is for $\tau = 10^8$~yr.}

\figcaption{The predicted time evolution of the various radial
        forces acting on fluxoids (per unit length) in a
        solitary neutron star, according to the A1 model and
        corresponding to the results in Fig.~1 and Fig.~2a. 
        The three curves represent the pinning force ${\rm F}_{\rm n}$,
        the drag force ${\rm F}_{\rm v}$, and the sum of the
        buoyancy and curvature forces ${\rm F}_{\rm b}+{\rm F}_{\rm c}$.}

\figcaption{Evolutionary tracks on the $B-P$ diagram for 
	solitary pulsars born with the different assumed initial field
	strengths, as predicted in the different models discussed in
        the text. The results shown are for the SIF and the DCC models, 
	while the latter is similar to those of the other FBE models.
	Positions of the neutron stars at various ages are marked on
	each track, and the spin-up line and the death line are also
        shown in each panel. Initial values of $B_{\rm c} = 0.9 B_{\rm s}$,
        and a value of $\tau = 10^7$~yr have been used.}

\figcaption{Similar graphs as in Figs~1 and 2, but for the tentatively
        assumed case of a small spin-down torque acting on the neutron
        star (corresponding to the
	dipole radiation torque alone for a star with nearly parallel
        spin and magnetic axes). The two panels on the {\em left} are
        for an initial value of the spin period
        $P_{\rm s} = 0.01$~s, while the {\em right} panels are for the
        case with initial $P_{\rm s} = 2.0$~s. Notice that the
        velocity curves for ${\rm V}_{\rm n}$ and ${\rm V}_{\rm p}$
        on the {\em left} panel coincide throughout.}

\figcaption{The predicted radio-active lifetimes of single pulsars
        versus their initial surface fields, according to the
        different field decay models discussed in the text, while
        ``Exp.'' represents the pure exponential decay model.
        Results for two different values of $\tau$ are shown
        in the two panels, as indicated.}

\figcaption{Final values of the surface magnetic field
        strengths of neutron stars evolved in low-mass binaries 
        versus initial orbital periods, as predicted in two of
        the FBE models. The other models (namely A2, B1, and B2)
        also produce similar results. The different curves, in
        each panel, correspond to the different assumed values of
        the companion mass-loss rate $\dot M_2$, as indicated by
        their logarithmic values, in units of $M_{\sun}$~yr$^{-1}$. Encircled dots
        represent observed binary radio pulsars that are descendants of
        wide low-mass binaries for which the initial orbital periods
	can be estimated.  Initial values of $B_{\rm s}=3.16 \times
        10^{12}$~G, $B_{\rm c} = 0.9 B_{\rm s}$, $P_{\rm s}= 0.1$~s,
        together with values of $\tau = 10^7$~yr, and $\eta=10$ have
        been used.}

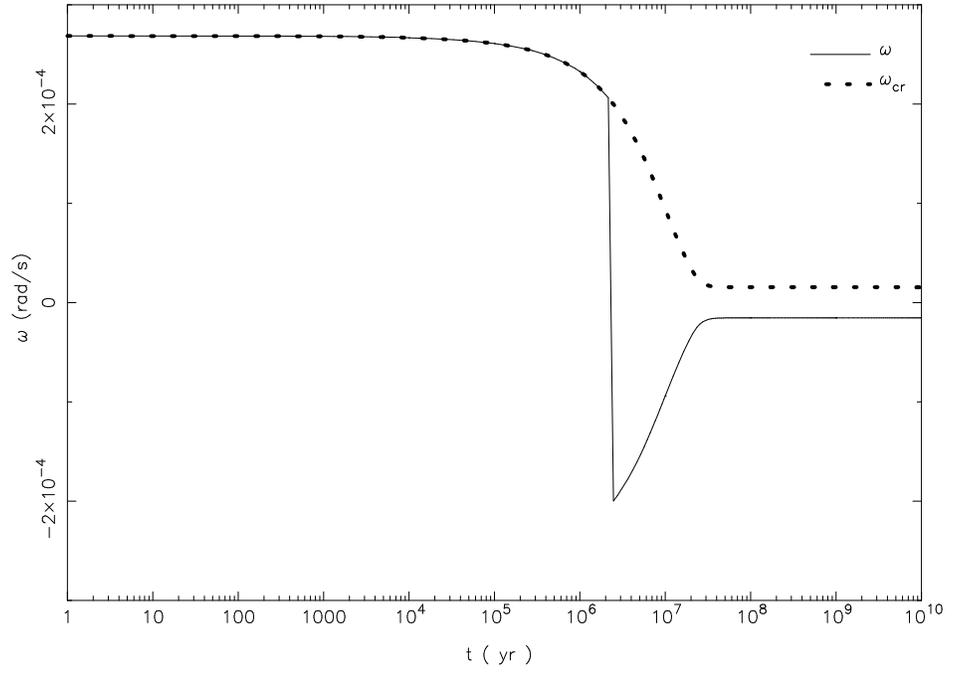
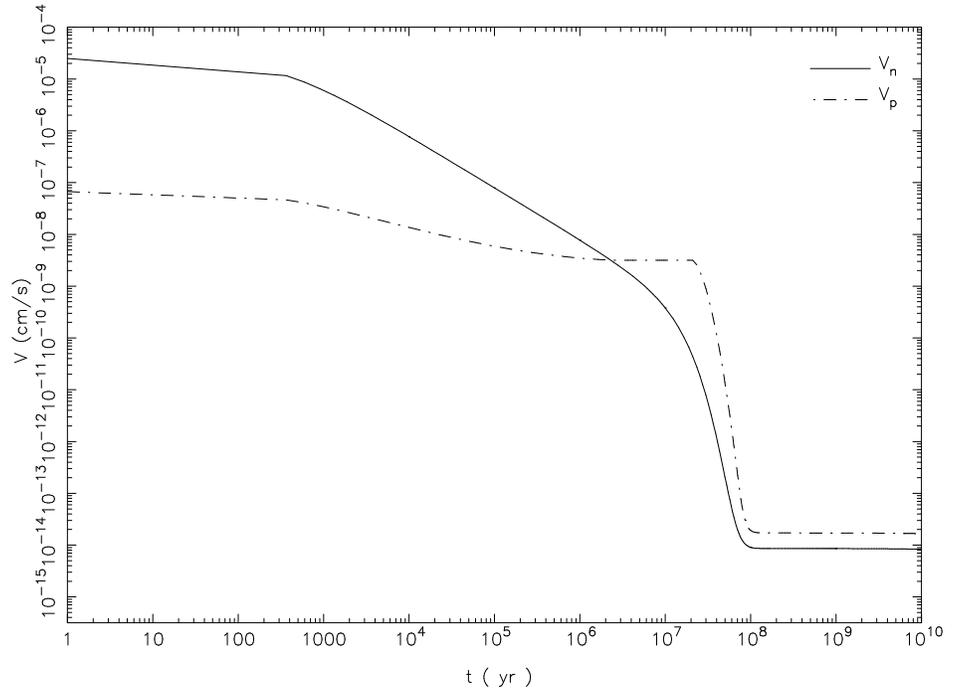
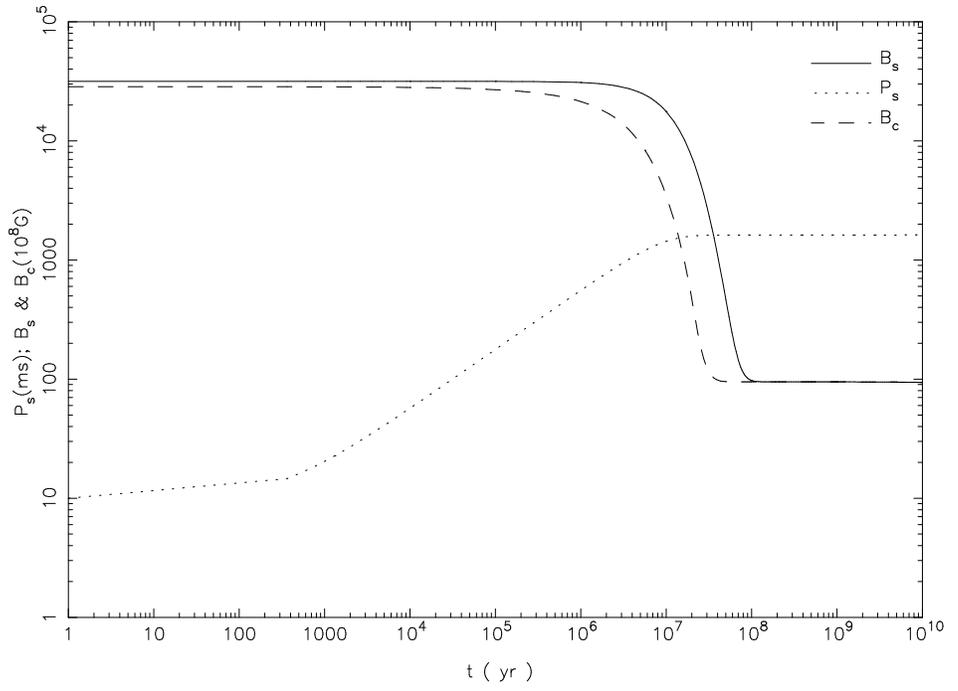
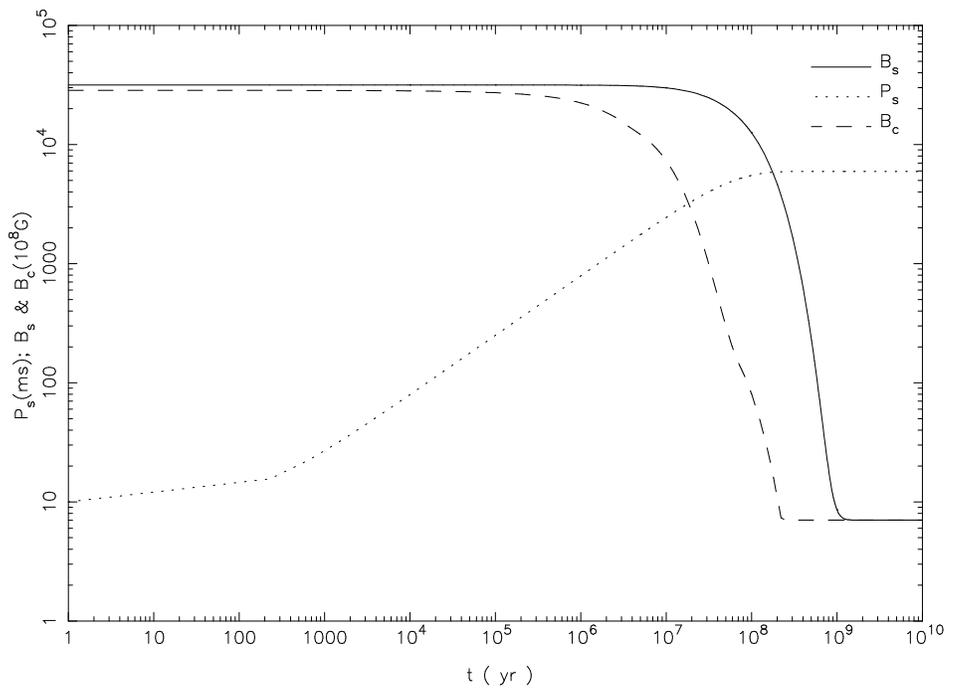
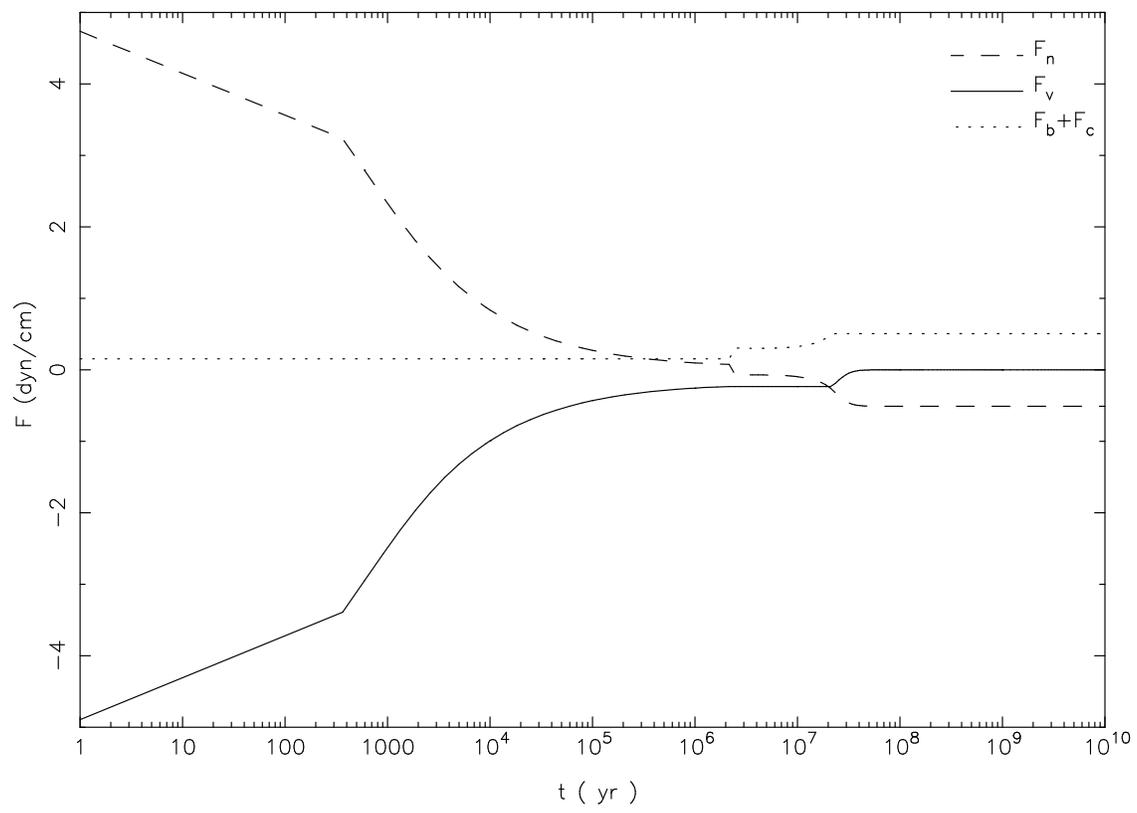
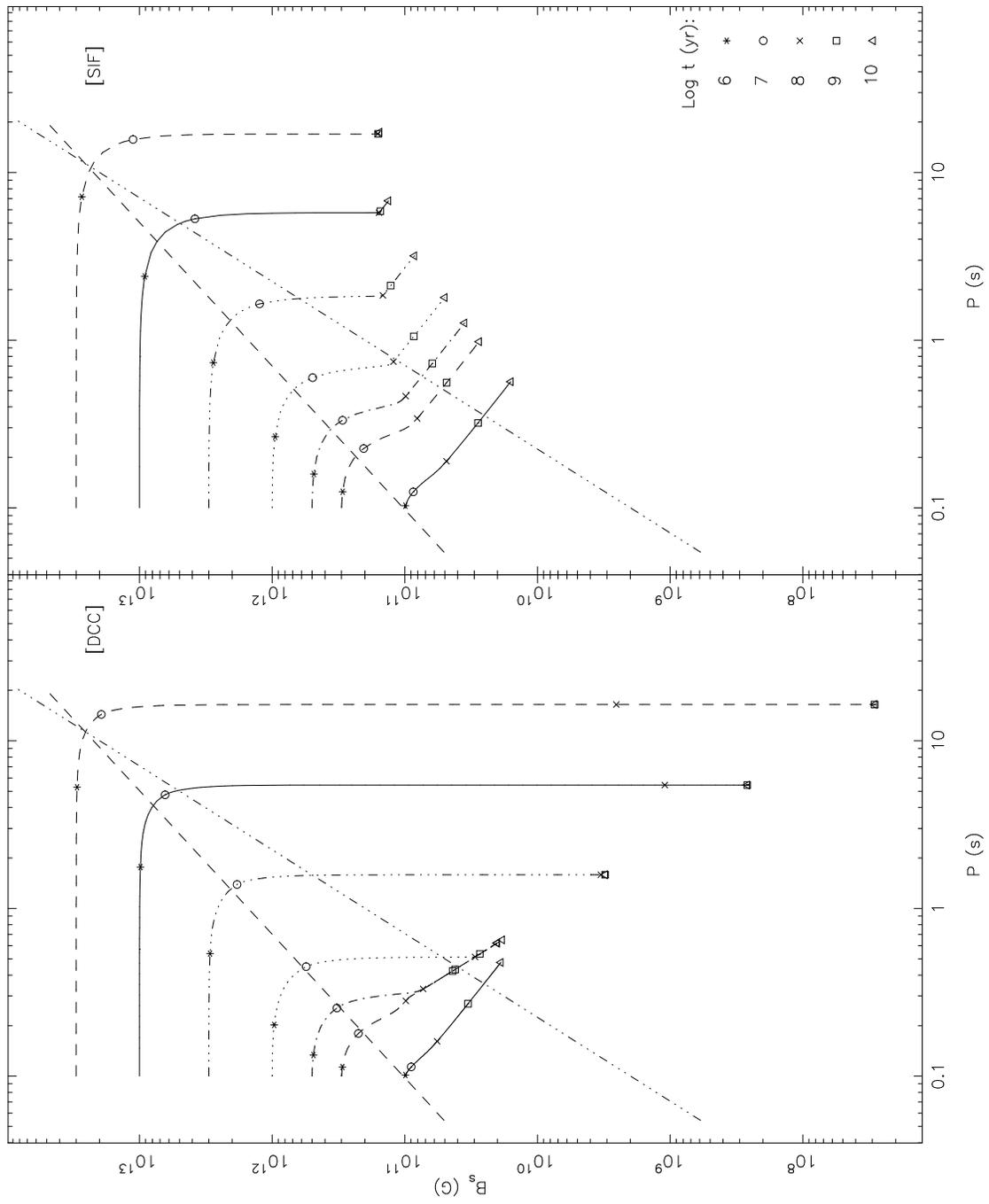
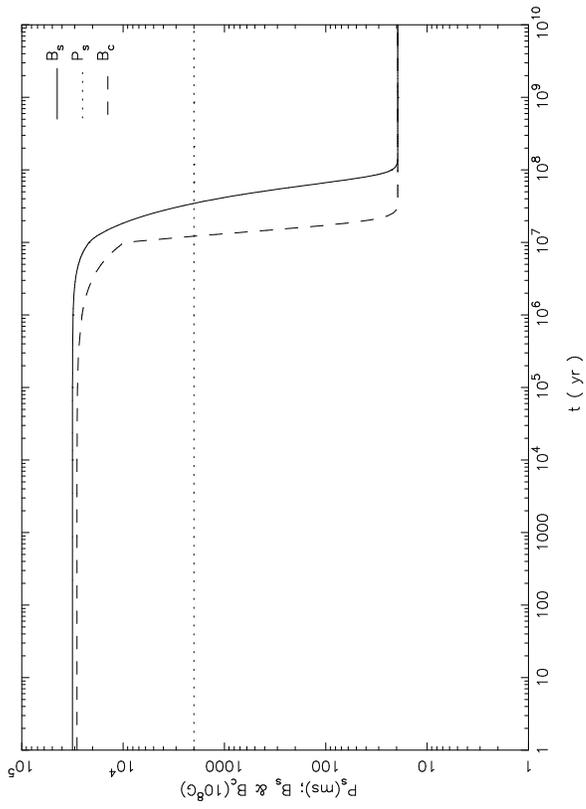
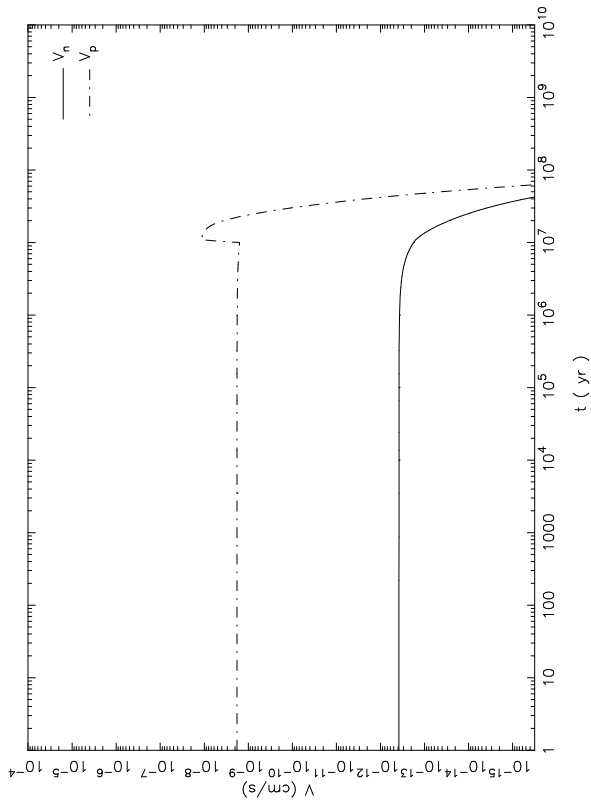
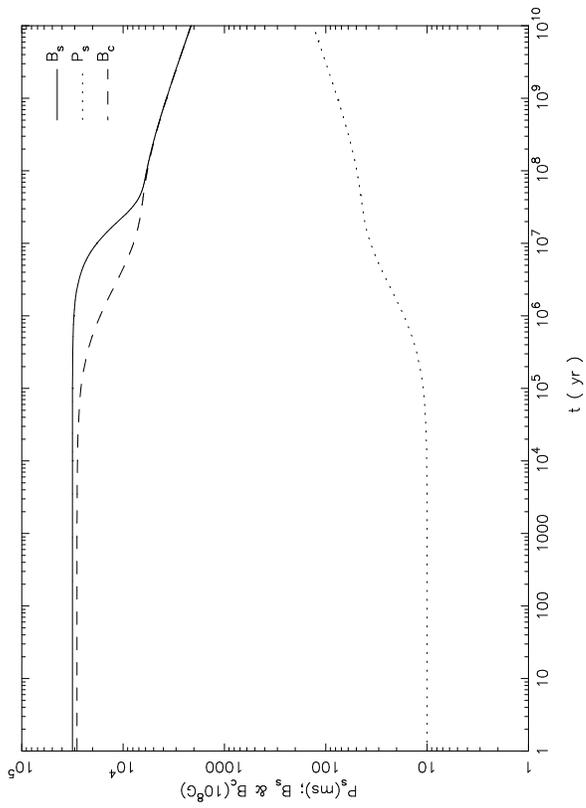
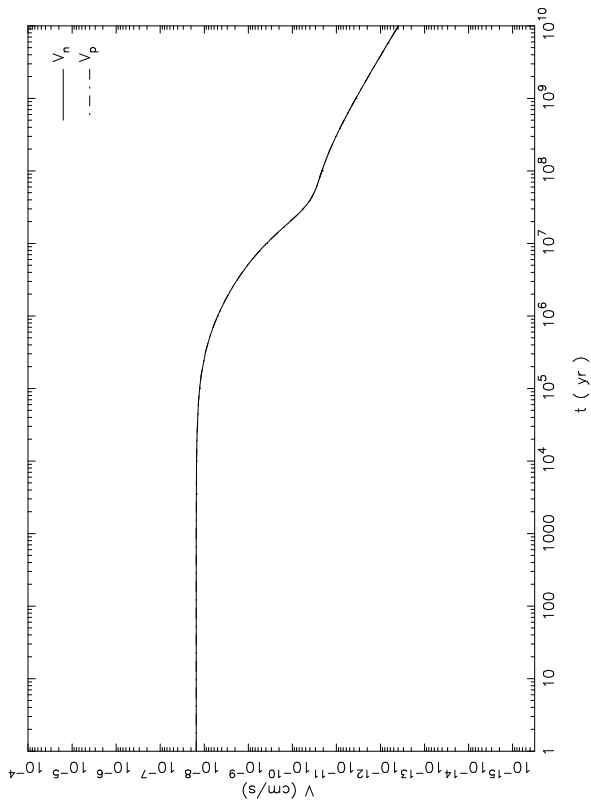
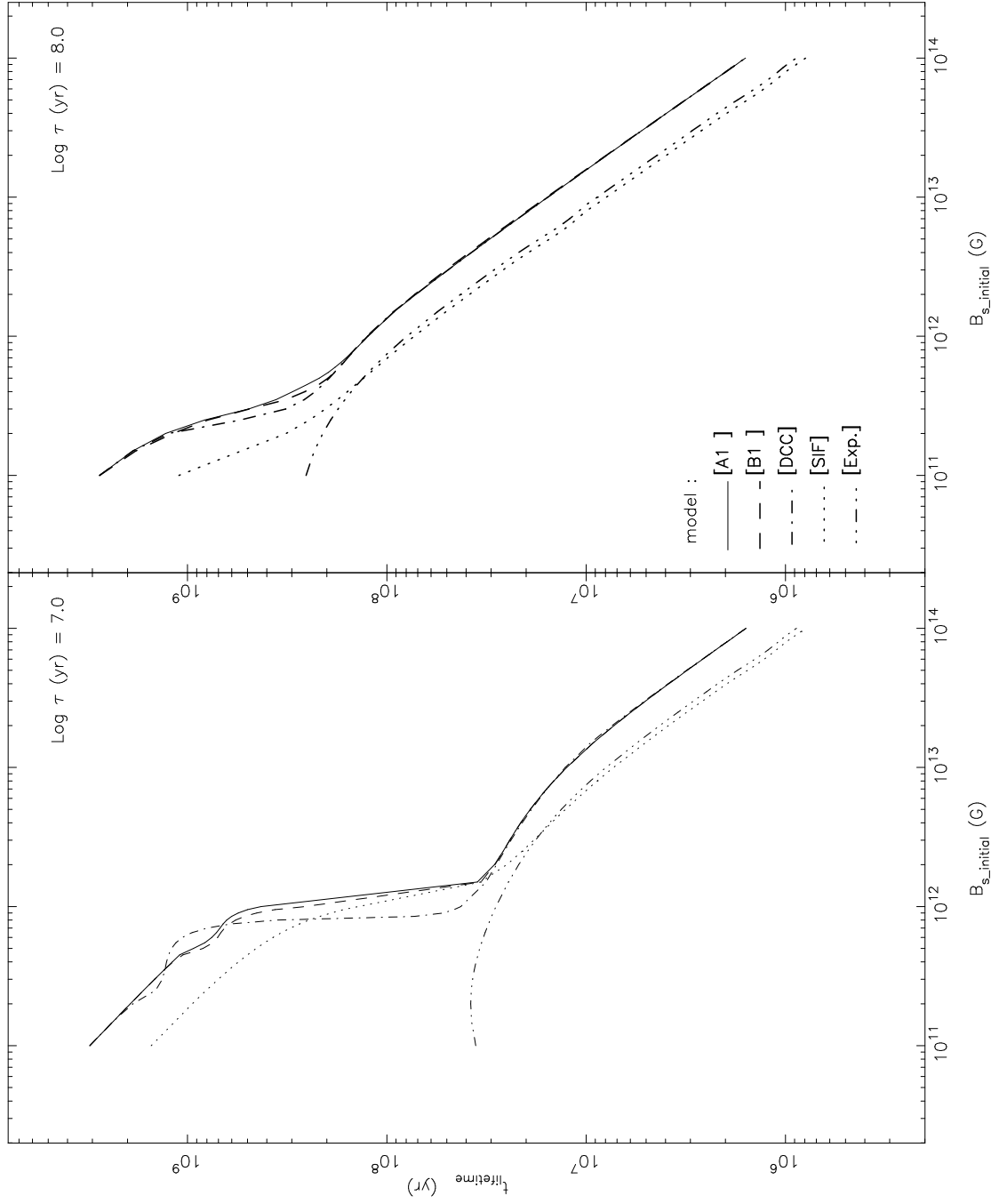
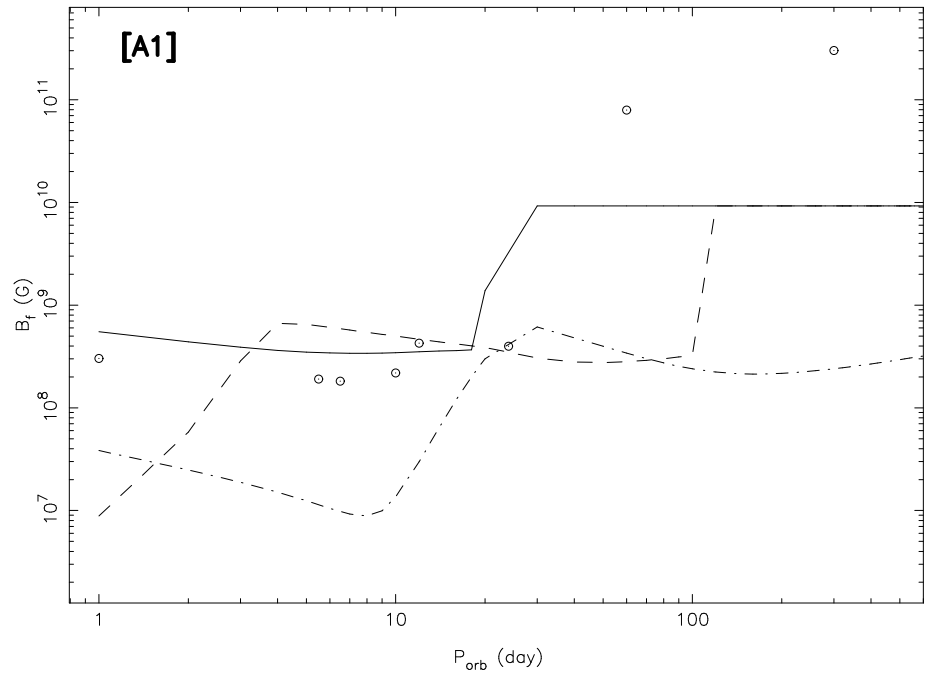
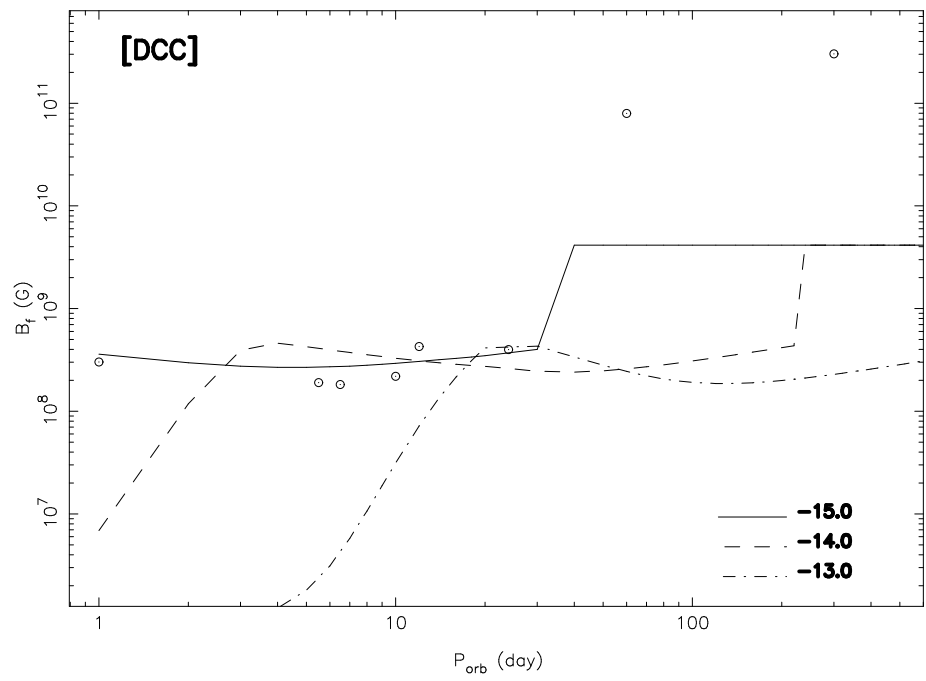
\end{document}